\documentclass[fleqn,usenatbib]{mnras}

\usepackage{orcidlink}
\usepackage{newtxtext,newtxmath}

\usepackage[thinc]{esdiff}
\usepackage{chemformula}

\usepackage[T1]{fontenc}

\DeclareRobustCommand{\VAN}[3]{#2}
\let\VANthebibliography\thebibliography
\def\thebibliography{\DeclareRobustCommand{\VAN}[3]{##3}\VANthebibliography}

\usepackage{graphicx}	
\usepackage{amsmath}	

\newcommand{\new}[1]{#1}
\newcommand{\newer}[1]{#1}
\newcommand{\newest}[1]{#1}

\newcommand{\newerer}[1]{#1}

\title[Optical transmission spectrum of HAT-P-44b]{LRG-BEASTS: Detection of sodium and evidence for water absorption in the hot Saturn HAT-P-44b}

\author[A. B. Claringbold et al.]{Alastair B. Claringbold$^{\orcidlink{0000-0003-1309-5558},1,2}$\thanks{E-mail: alastair.claringbold@warwick.ac.uk},
Peter J. Wheatley$^{\orcidlink{0000-0003-1452-2240},1,2}$\thanks{E-mail: p.j.wheatley@warwick.ac.uk},
James Kirk$^{\orcidlink{0000-0002-4207-6615},3}$,
Eva-Maria Ahrer$^{\orcidlink{0000-0003-0973-8426},4,1,2}$,
Ian Skillen$^{\orcidlink{0000-0002-3849-8276},5}$,
\newauthor
Matteo Brogi$^{\orcidlink{0000-0002-7704-0153},6,7}$,
George W. King$^{\orcidlink{0000-0002-3641-6636},8}$,
James McCormac$^{\orcidlink{0000-0003-1631-4170},1,2}$
\\
$^{1}$Department of Physics, University of Warwick, Gibbet Hill Road, Coventry CV4 7AL, UK\\
$^{2}$Centre for Exoplanets and Habitability, University of Warwick, Gibbet Hill Road, Coventry CV4 7AL, UK\\
$^{3}$Department of Physics, Imperial College London, Prince Consort Road, London SW7 2BW, UK\\
$^{4}$Max Planck Institute for Astronomy, Königstuhl 17, 69117 Heidelberg, Germany\\
$^{5}$Isaac Newton Group of Telescopes, Apartado de correos 321, E-38700 Santa Cruz de La Palma, Canary Islands, Spain\\
$^{6}$Dipartimento di Fisica, Universit\'a degli Studi di Torino, via Pietro Giuria 1, I-10125, Torino, Italy \\
$^{7}$INAF-Osservatorio Astrofisico di Torino, Via Osservatorio 20, I-10025 Pino Torinese, Italy \\
$^{8}$Department of Astronomy, University of Michigan, Ann Arbor, MI 48109, USA
}

\date{Accepted XXX. Received YYY; in original form ZZZ}

\pubyear{2024}

\begin{document}
\label{firstpage}
\pagerange{\pageref{firstpage}--\pageref{lastpage}}
\maketitle

\begin{abstract}

\noindent We present the low-resolution optical transmission spectrum of \newer{the inflated hot Saturn HAT-P-44b}. The planet is a close sibling in radius (1.24 $\mathrm{R_{Jup}}$), temperature (1100 K), and mass (0.35 $\mathrm{M_{Jup}}$) to the exceedingly well-characterized WASP-39b. 
Using the ACAM instrument on the William Herschel Telescope (WHT), we obtain a \new{transmission spectrum} with sub-scale height precision of \new{246} ppm, with a wavelength range of 495 -- 874 nm and a 20 nm resolution, despite a relatively faint host star ($V\mathrm{_{mag} = 13.2}$). We detect absorption due to sodium with \new{ 3.9$\sigma$ confidence. Atmospheric retrieval of the transmission spectrum also reveals \newer{evidence for} \ch{H2O} absorption and Rayleigh scattering from \ch{H2} gas consistent with a cool 800 K atmosphere and a super-solar metallicity of $7${\raisebox{0.5ex}{\tiny$\substack{+16 \\ -5}$}}$\times$solar. Comparison of retrieval models disfavour the inclusion of a \newer{super-Rayleigh scattering slope or high-altitude clouds (at $<1$ mbar) while being agnostic towards the presence of mid-altitude clouds}. Our transmission spectrum \newer{of HAT-P-44b shows strong similarity} to that of its sibling WASP-39b.} This is the tenth planet in the LRG-BEASTS (Low-Resolution Ground-Based Exoplanet Atmosphere Survey using Transmission Spectroscopy) survey.
\end{abstract}




\section{Introduction}

Transmission spectroscopy from both space- and ground-based telescopes is the premier tool for the characterization of exoplanet atmospheres \citep{charbonneau2002detection,redfield2008sodium,snellen2010orbital,kreidberg2014precise,jwst2023identification}. With deep transit depths and large scale heights, hot Jupiters provide the best signal to noise ratio spectra, and therefore enable the most detailed insights into exoplanet atmospheres.
Transmission spectroscopy across the hot Jupiter population has revealed considerable diversity in both chemical and aerosol properties, with evidence for molecular absorption \citep[e.g.,][]{de2013detection,deming2013infrared,wakeford2017complete,alderson2023early}, atomic species \citep[e.g.,][]{redfield2008sodium,sing2011gran,alderson2020lrg,alam2021evidence,ahrer2022lrg,feinstein2023early}, Rayleigh and super-Rayleigh scattering slopes \citep[e.g.,][]{des2008rayleigh,sing2015hst,kirk2017rayleigh,alderson2020lrg}, muted features taken as evidence for high-altitude clouds \citep[e.g.][]{kreidberg2014clouds,lendl2016fors2,spyratos2021transmission,ahrer2023lrg}, as well as spectroscopic impacts of stellar activity \citep[e.g.,][]{oshagh2014impact,mccullough2014water,rackham2017access}.

Near infrared transmission spectroscopy at both low and high resolution has unlocked the chemical composition of many exoplanets, with the improvements in precision and wavelength coverage offered by JWST enabling especially detailed analyses \citep[e.g.,][]{fu2022water,radica2023awesome,rustamkulov2023early}. The deduction of elemental ratios from chemical abundances can in principle be used to understand formation and migration processes of the planets \citep[e.g.,][]{oberg2011effects,helling2014disk,madhusudhan2014toward,booth2017chemical,espinoza2017metal}. The detection of some atmospheric species out of equilibrium can provide direct evidence of chemical processes including photochemistry, like the detection of \ch{SO2} in WASP-39b \citep{alderson2023early,rustamkulov2023early,tsai2023photochemically}. Mid-infrared transmission spectroscopy has been able to identify spectral features of the cloud species 
\citep{grant2023jwst,dyrek2024so2}.

Atmospheric models used to interpret transmission spectra can contain considerable degeneracies between chemical abundances, scale height, and aerosol parameters. Modelling of cloud and haze formation processes gives hints to the atmospheric species responsible for aerosols as well as trends in planetary populations \citep[e.g.][]{ackerman2001precipitating,heng2016cloudiness,stevenson2016quantifying,kawashima2019theoretical,gao2020aerosol,gao2021aerosols}.

\new{The size and location of aerosol species can be best constrained by observations at optical wavelengths and low spectral resolution,} through the identification of scattering slopes and the muting of the wings or cores of sodium and potassium absorption.
Strong Rayleigh or super-Rayleigh scattering slopes in optical spectra indicate the presence of small aerosol particles \citep[e.g.,][]{des2008rayleigh,kirk2017rayleigh,alderson2020lrg}. Flat, featureless optical spectra are attributed to larger aerosol particles \citep[e.g.,][]{kreidberg2014clouds,lendl2016fors2,ahrer2023lrg}. A lack of aerosols in the atmosphere can be inferred from the detection of a pressure-broadened sodium feature \citep{seager2000theoretical}, as seen in WASP-39b, WASP-96b, and WASP-62b \citep[][]{fischer2016hst,nikolov2018absolute,alam2021evidence}.
Optical transmission spectroscopy is therefore a powerful tool in both reducing degeneracies in chemical composition and distinguishing between clear, hazy, and cloudy atmospheres.

In order to investigate the diversity of clouds and hazes across the hot gaseous exoplanet parameter space, we require a large sample size. This has the added benefit of providing optical transmission spectra that can both synergize with space-based infrared observations, as well as inform target selection for science cases which require the high precision detections facilitated by clear atmospheres. The Low Resolution Ground-Based Exoplanet Atmosphere Survey using Transmission Spectroscopy (LRG-BEASTS) is designed to provide a large sample of hot Jupiters with homogeneously analysed optical transmission spectra. LRG-BEASTS has demonstrated the ability to achieve comparable precision on 4m-class telescopes like the William Herschel Telescope (WHT) and New Technology Telescope (NTT) to that obtained by 10-m class telescopes and the Hubble Space Telescope (HST).

Previous characterizations with LRG-BEASTS have detected hazes in HAT-P-18b \citep{kirk2017rayleigh} and WASP-80b \citep{kirk2018lrg}, grey clouds in WASP-52b \citep{louden2017precise} and HATS-46b \citep{ahrer2023lrg}, and sodium absorption and super-Rayleigh slopes in WASP-21b \citep{alderson2020lrg} and WASP-94 Ab \citep{ahrer2023lrg}. Combined analyses involving LRG-BEASTS data have aided in the interpretation of the atmospheres of WASP-39b \citep{kirk2019lrg}, WASP-103b \citep{kirk2021access}, WASP-25b \citep{mcgruder2023access} and WASP-94Ab \citep{ahrer2024atmospheric}. These join results from other ground-based surveys using the Gran Telescopio Canarias \citep[e.g.][]{parviainen2016gtc,murgas2017gtc}, Magellan/IMACS \citep[ACCESS, e.g.][]{rackham2017access,espinoza2019access}, Gemini/GMOS \citep[e.g.][]{huitson2017gemini,panwar2022new}, and VLT/FORS2 \citep[e.g.][]{nikolov2016vlt,sedaghati2016potassium,spyratos2021transmission}.

In this paper, we use the ACAM instrument on the WHT to present the transmission spectrum of the inflated hot Saturn HAT-P-44b \citep{hartman2014hat}. With a radius of $1.24\substack{+0.11 \\ -0.05}$\, $\mathrm{R_{Jup}}$, mass of $0.35\pm0.03$$\mathrm{M_{Jup}}$, and equilibrium temperature of $1110\substack{+50 \\ -30}$\, K, HAT-P-44b is a close sibling to the well-characterized WASP-39b (the target of the JWST Early Release Science programme as well as a previous LRG-BEASTS target), which has a comparable radius of $1.27\pm0.04$ $\mathrm{R_{Jup}}$, mass of $0.281\pm0.032$ $\mathrm{M_{Jup}}$, and equilibrium temperature of $1166\pm14$ K. The deep transit depth, low gravity, and high equilibrium temperature of HAT-P-44b gives it a strong expected signal of 288 ppm for a single scale height. \newer{HAT-P-44b has been observed using the Exoplanet Transmission Spectroscopy Imager on the McDonald Observatory Otto Struve 2.1m telescope, with a reconnaissance spectrum obtained in eight wavelength bins \citep{oelkers2025ground}. The variance of transit depth across the eight bins is used to suggest that HAT-P-44b may have a clear atmosphere.}

We describe our observations in Section \ref{sec:observations}, and our reduction and analysis in Sections \ref{sec:reduction} and \ref{sec:fitting}. We conduct an atmospheric retrieval analysis to interpret our transmission spectra in Section \ref{sec:retrieval}, and present our discussion and conclusions in Section \ref{sec:discussion} and Section \ref{sec:conclusion}.

\section{Observations}
\label{sec:observations}

\begin{figure}
    \centering
    \includegraphics[width=0.48\textwidth]{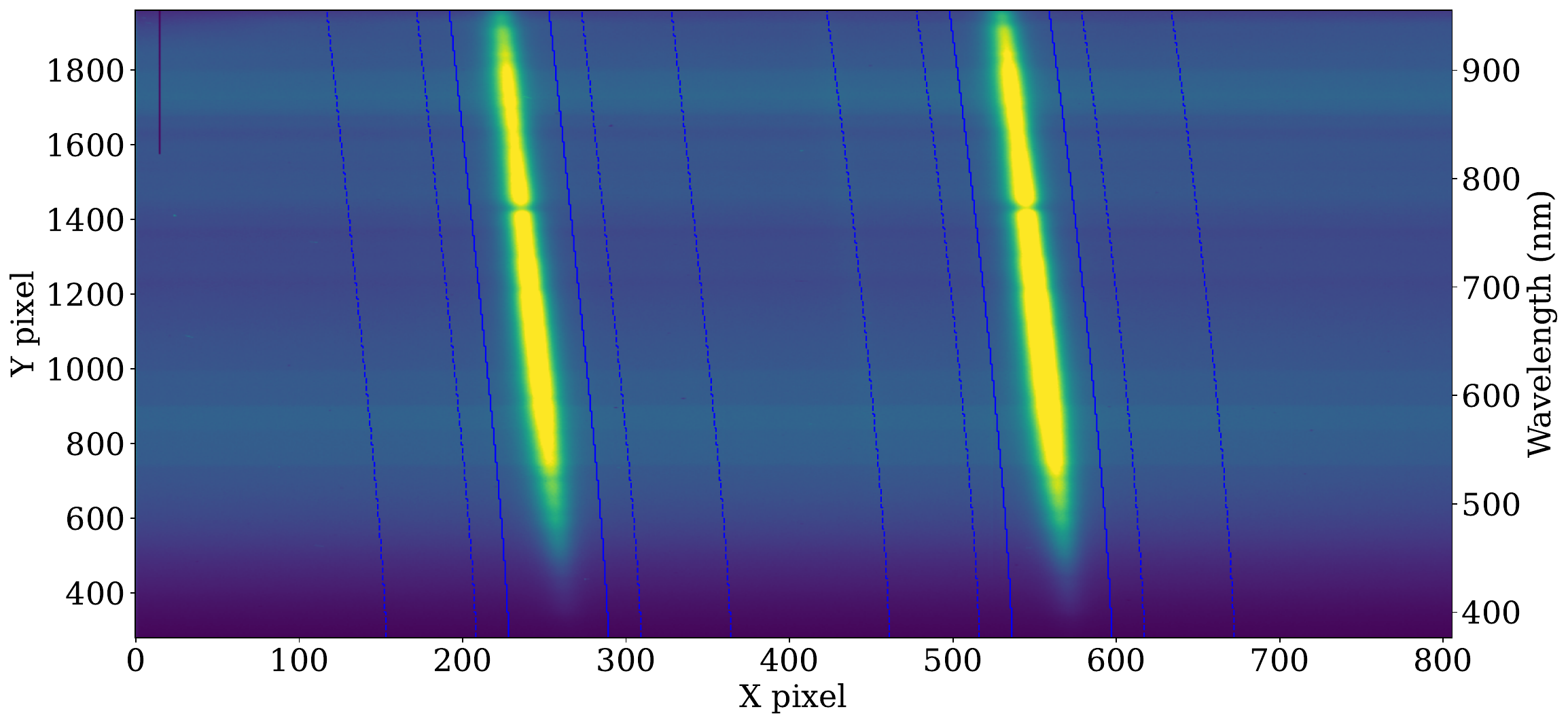}
    
    \includegraphics[width=0.48\textwidth]{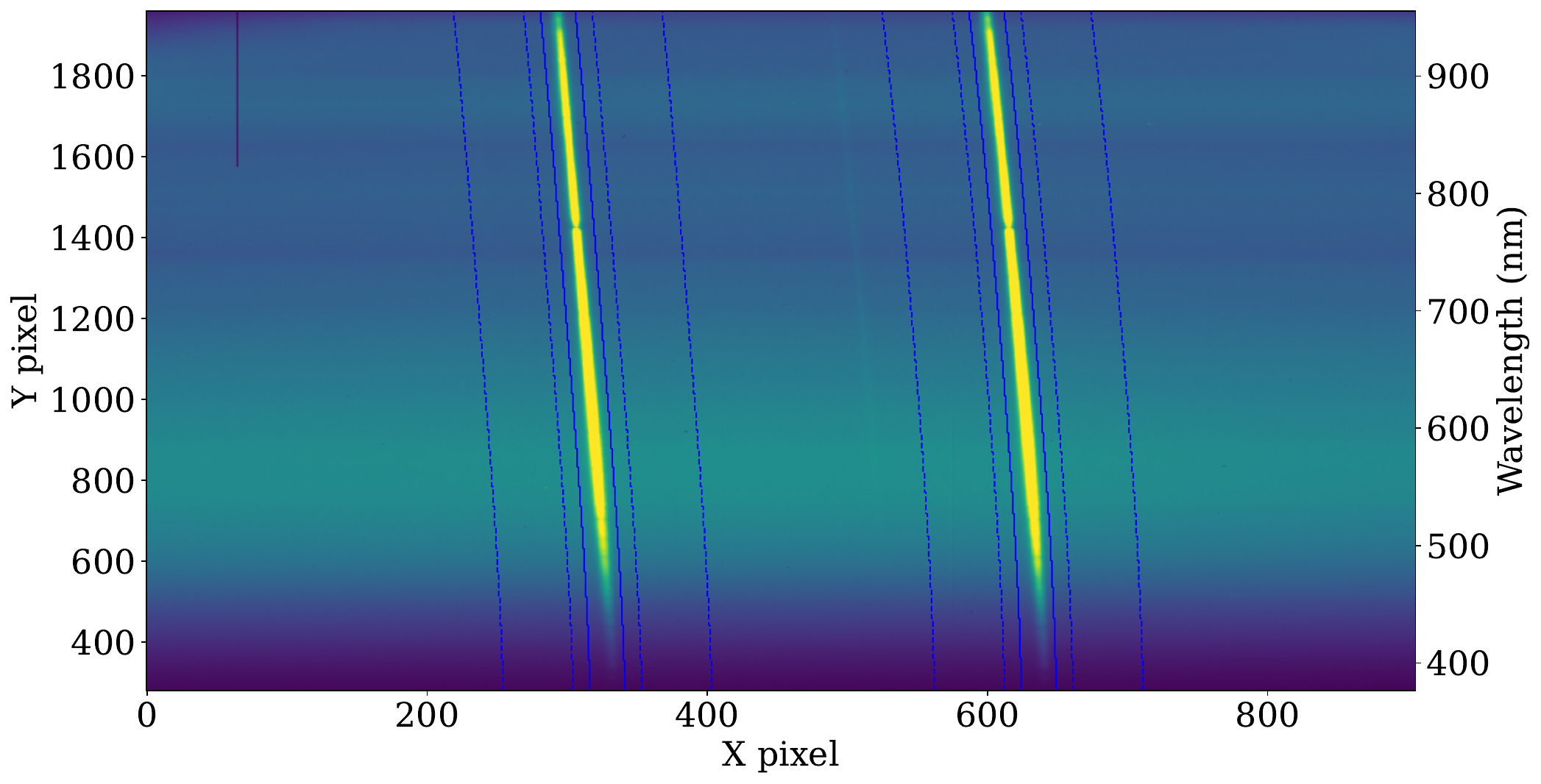}
    
    \includegraphics[width=0.48\textwidth]{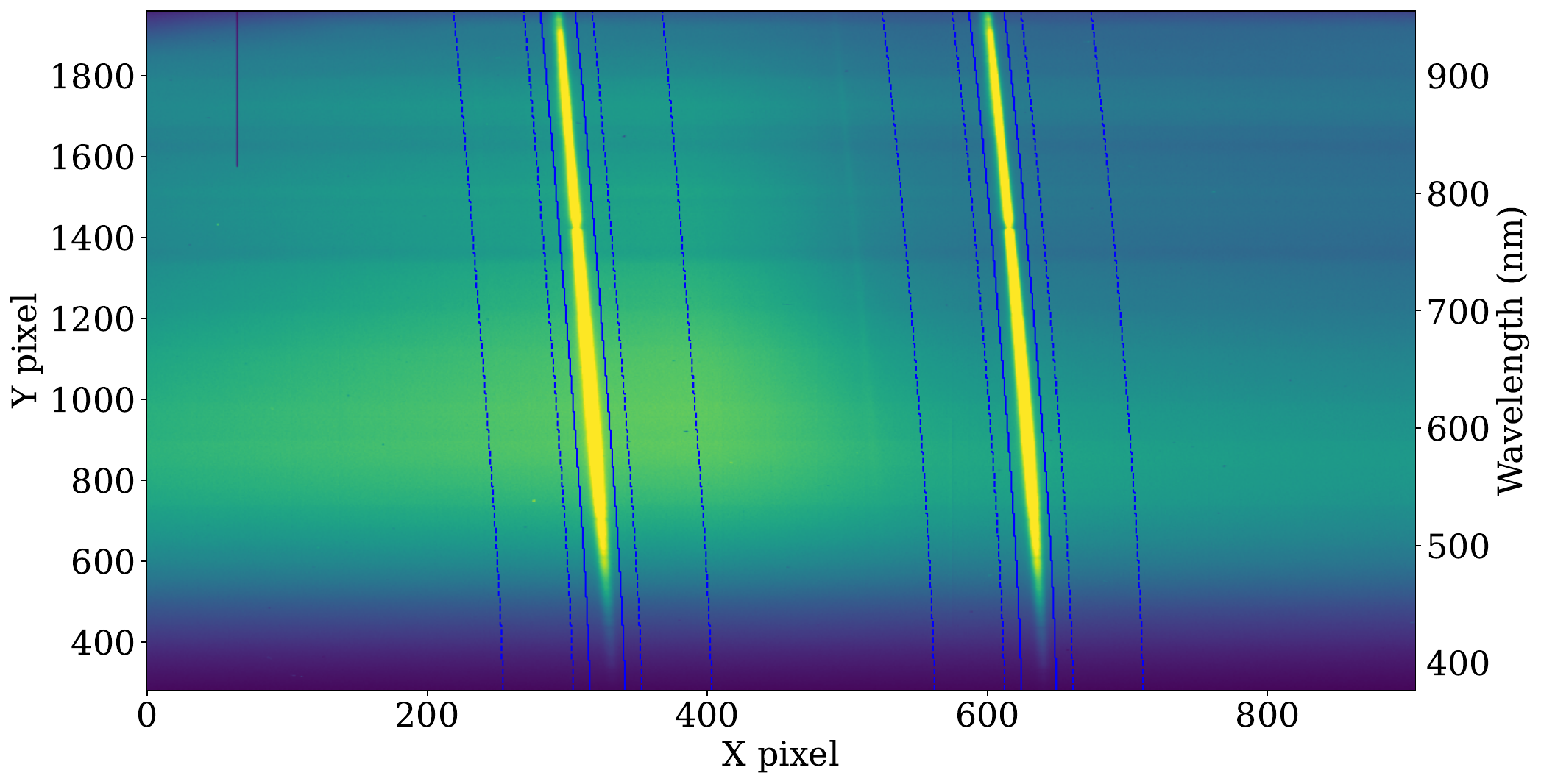}

    \includegraphics[width=0.48\textwidth]{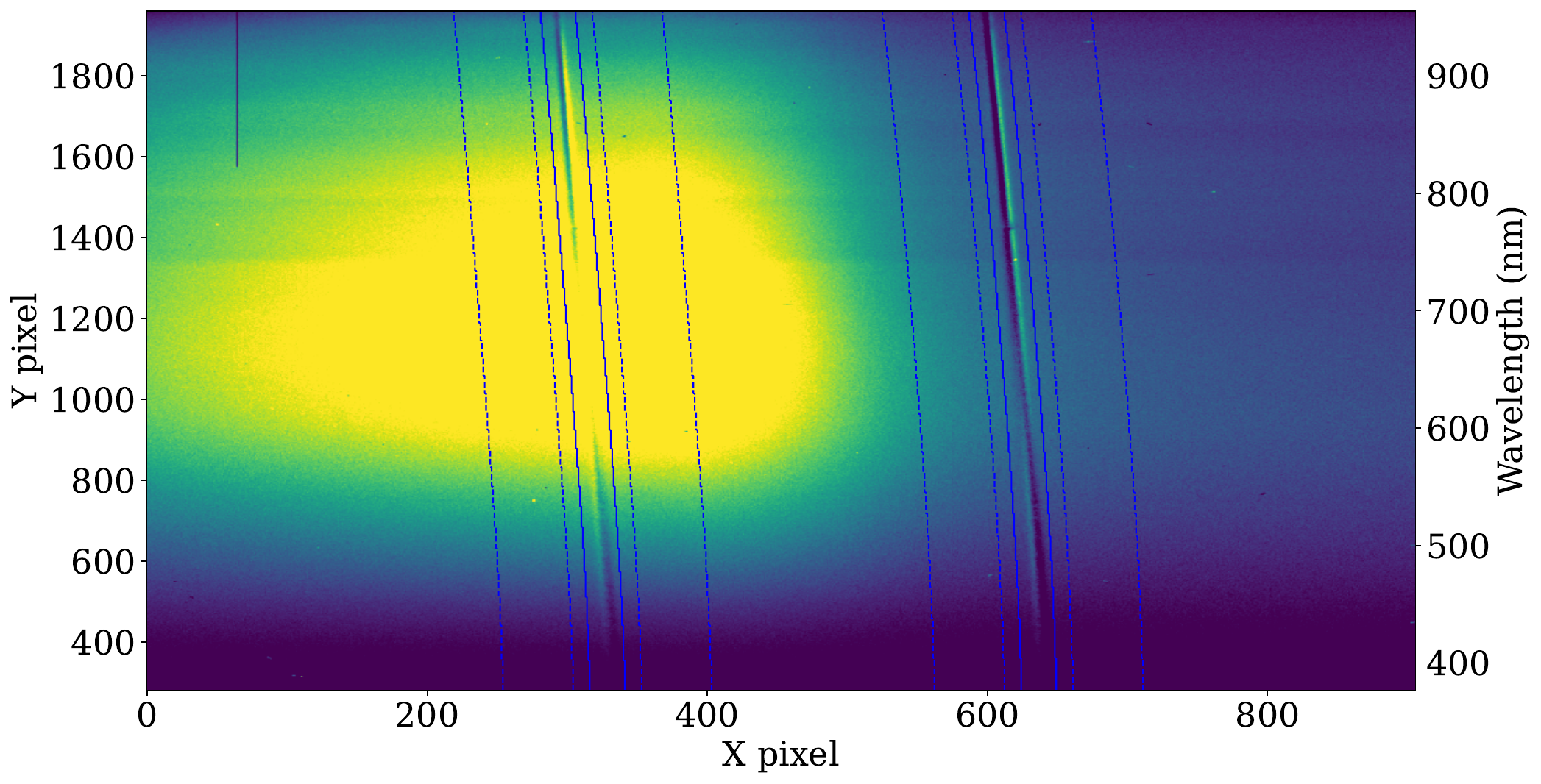}
    \caption{\new{Example ACAM frames demonstrating the used extraction (solid blue lines) and background (dashed blue lines) regions from the first night (first panel) and second night (second panel). We also observed a strong 2D enhancement on the detector in the second night for some of the transit, depicted in the third panel, and highlighted with a difference frame between without-scattering and with-scattering frames with similar FWHM for the second night in the final panel. \newerer{The approximate wavelength of each Y pixel is indicated.} Each X pixel spans 0.25 arcseconds.}}
    \label{fig:extraction-frames}
\end{figure}

We observed two transits of HAT-P-44b on the nights of 2018 March 26 and 2019 March 18 using the ACAM instrument \citep{benn2008acam} on the WHT, providing low resolution spectroscopy over a wide wavelength range ($\sim$ 350 -- 940 nm) with a wide field of view (8 arcmin). We used fast readout mode, and did not employ an order-blocking filter for either night. This is the same instrument setup for long-slit spectroscopy as \citet{kirk2017rayleigh,kirk2018lrg,kirk2019lrg} and \citet{alderson2020lrg}. \new{The VPH grism dispersion is 0.33 nm/pixel.}

We used a 7.6 arcmin long, 40 arcsec wide slit \newer{rotated to a position angle of 80.9 degrees to observe the target and a comparison star} while avoiding differential slit losses. We used the same comparison, 2MASS J14124195+4701058, located 1.28 arcmin from HAT-P-44 
for both nights. The comparison is both a good brightness ($\Delta V = 0.21$) and colour match ($\Delta (B-V) = 0.11$), and therefore enables effective differential spectroscopy. 

For the first night, we used a long exposure time of 300 s due to the relative faintness of HAT-P-44 ($V\mathrm{_{mag} = 13.2}$). The moon was at 77\% illumination and 72 degrees from the target, setting during the post-transit observation. The first hour of the observation was affected by clouds and poor seeing ($4"$), and seeing remained high throughout the night ($1.5-3"$). There is a small gap after the fifth exposure, as guiding briefly failed due to clouds. \new{The effective spectral resolution of this night is 12 nm pre-transit, and around 5 nm for the rest of the night, once the seeing had improved.}

\begin{figure}
    \centering
    \includegraphics[width=0.49\textwidth]{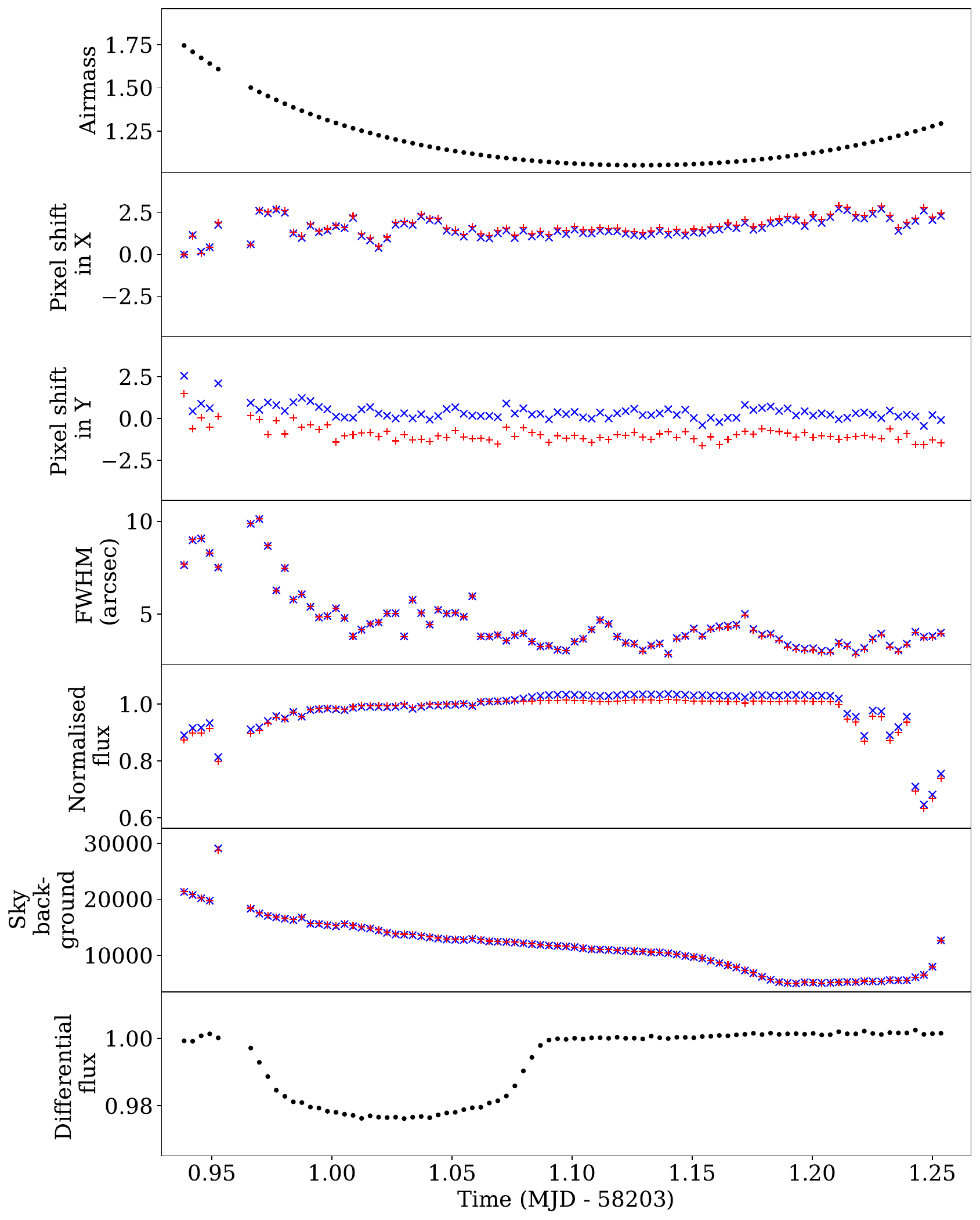}
    \caption{Ancillary data for the first night of observation of HAT-P-44b. The blue crosses correspond to HAT-P-44b and the red crosses correspond to the comparison star. Each panel represents the variation of various quantities throughout the night, from top to bottom: target airmass, the shift in trace x and y centroid across the slit, the full width half maximum (FWHM) of the stellar lines, the normalized flux, the total sky background counts, and the white-light light-curve differential flux. The impact of clouds is evident as dips in the normalized flux in the fifth and final twelve exposures. The poor seeing at the beginning of the night is apparent in the FWHM plot, and the kink in the sky background curve near the end of the night is due to moonset.}
    \label{fig:ancillary_plots}
\end{figure}

The second night of observation took place during nearly full moon (94\% illumination at 60 degrees from the target) and had better seeing ($1.3"$), so we chose a reduced exposure time of 180 s. Clouds affected 5 of the pre-transit frames, causing a dip in the flux and spike in the background. During the transit we observed a sudden enhancement in the background localised to the blue end of the target spectrum that lasted for $\sim$40 minutes, \new{which can be seen in Fig. \ref{fig:extraction-frames}}. We tentatively attribute this to indirect scattered moonlight,
given its abrupt appearance and disappearance and the presence of the full moon. 
\new{The effective resolution for the second night is 4\,nm.}

\begin{figure}
    \centering
    \includegraphics[width=0.49\textwidth]{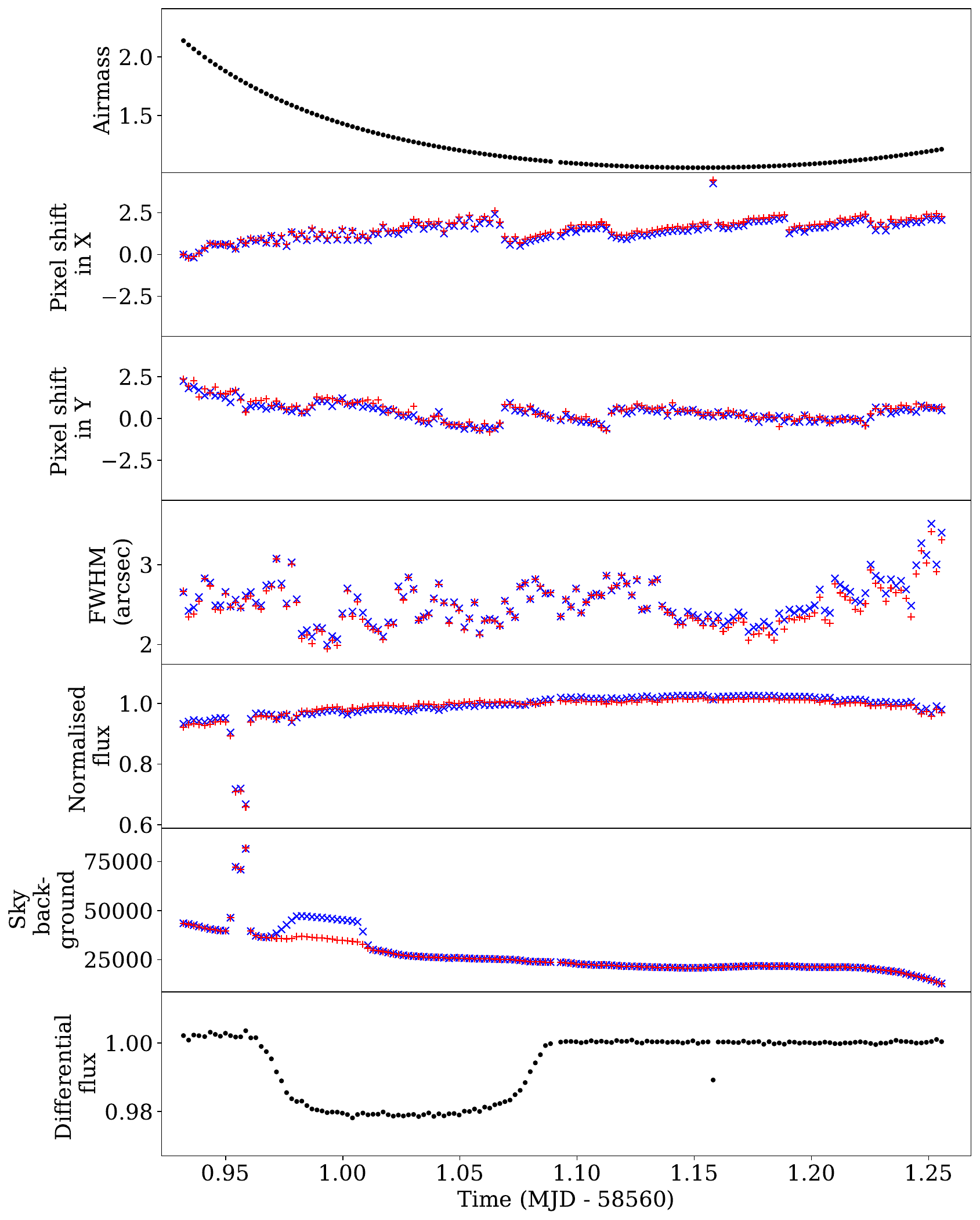}
    \caption{Ancillary data for the second night of observation of HAT-P-44b. The blue crosses correspond to HAT-P-44b and the red crosses correspond to the comparison star. From top to bottom: target airmass, the shift in trace centroid across the slit, the full width half maximum (FWHM) of the stellar lines, the total sky background counts, and the white-light light-curve differential flux. The minimum in normalized flux immediately before transit is due to clouds, while we believe differential enhancement in background to be indirect scattered moonlight due to its abrupt appearance and disappearance.}
    \label{fig:ancillary_plots2}
\end{figure}


Both observations used fast readout mode and a smaller-than-standard window to minimise \new{the overhead time to 6.4 s}. For both nights we also took lamp and sky flats, biases, and HeAr arc frames. We did not use any flat frames in our final data reduction as we found it increased the noise in our data, in line with previous analyses from LRG-BEASTS \citep[e.g.,][]{kirk2017rayleigh,alderson2020lrg,ahrer2023lrg}, ACCESS \citep[e.g.][]{rackham2017access}, \newer{ VLT/FORS2 \citep[e.g.][]{gibson2017vlt}, and Gemini \citep[e.g.][]{huitson2017gemini}. The use of flat-fielding in ACAM transmission spectroscopy was explored by \citet{KirkThesis}, who found it was not possible to construct a well-illuminated flat field, and the use of flat fielding introduced red noise across multiple datasets. In addition, as 1620-5400 pixels are included in each wavelength bin of our final transmission spectrum, the impact of pixel-to-pixel variations is negligible.} We constructed a median-combined master bias frame and subtracted it from each science frame. The airmass, x pixel shift, \new{y pixel shift}, trace full width at half maximum (FWHM), normalized flux, sky background flux, and differential flux for each night are presented in Fig. \ref{fig:ancillary_plots} and Fig. \ref{fig:ancillary_plots2} respectively. We applied manual guiding corrections in the second night based on real-time spectral analysis, which is apparent as discontinuities in x and y positions in Fig. \ref{fig:ancillary_plots2}.

\section{Data Reduction}
\label{sec:reduction}

We use the {\tt Tiberius} pipeline for spectral extraction and light-curve fitting introduced in \citet{kirk2017rayleigh,kirk2021access}, as is standard for LRG-BEASTS data \citep[e.g.][]{kirk2018lrg,kirk2019lrg,ahrer2022lrg}, and has also been used for JWST observations \citep[e.g.][]{alderson2023early,rustamkulov2023early,moran2023high,kirk2024jwst}. Our analysis closely follows that of \citet{alderson2020lrg}, with the key steps summarized below.

To extract the traces of HAT-P-44 and the comparison star, we fit the position of the trace with a quartic function in the dispersion direction and a Gaussian for the trace width, and place an aperture over it. To remove the sky background we subtract a polynomial fitted to background regions on either side of the trace on a row-by-row basis. We clip outliers in the sky background of three or more standard deviations and then refit the polynomial.We iterate the aperture width and background region and polynomial order, to minimise the rms of the residuals in the white-light light-curves. For the first night we use a wide aperture of 60 pixels ($\sim$15 arcsecs) \new{full-width}, with a large 20 pixel offset to a 55 pixel background region on either side of the target spectrum. For the second night we use a narrower 24 pixel ($\sim$6 arcsecs) \new{full-width} aperture and a 50 pixel background region offset by 12 pixels. The aperture is much wider in the first night because the seeing was much greater\new{, and is a compromise between fitting the 40 pixel FWHM pre-transit, and the much lower 15 pixel FWHM for the rest of the night.} In both cases we find that a quadratic fit for background subtraction minimises the white-light light-curve rms. As the seeing was highly variable on the first night, the chosen width is a compromise value to avoid systematics associated with a time-varying aperture width.

After spectral extraction, we remove cosmic rays from the spectra using a running median to identify outliers, \new{replacing identified pixels with the mean of their neighbours. We also remove outliers directly from the original frame in the background region when performing background subtraction.} We then resample the spectra onto consistent pixels in wavelength by fitting Moffat profiles \newer{to ten stellar absorption features spanning from 486 to 866 nm} to calculate the pixel shifts of the spectrum \new{(including stretching and compression)}, which we then smooth \newest{in time} using a cubic function before resampling. \newest{We then use a single fit to 13 stellar and telluric lines from a reference spectrum in the middle of the night to assign a wavelength scale to the aligned spectra. We permit differences in radial velocities between the target and comparison star.}

\begin{figure}
    \centering
    \includegraphics[width=0.49\textwidth]{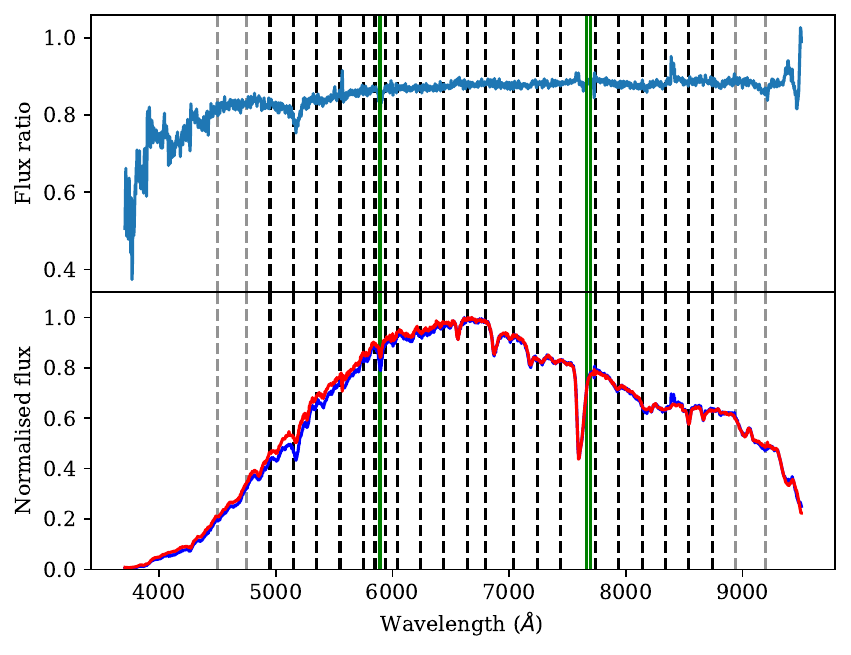}
    \caption{Flux ratio between the target and comparison (top panel), and normalized flux of the target (bottom panel, blue) and comparison (bottom panel, red) as a function of wavelength, for an example exposure. The wavelength bins used in the creation of our transmission spectra as displayed as black dotted lines, and the wavelengths of Na I and K I absorption are highlighted as green lines at 589 nm and 770 nm respectively. Additional bins discluded from our final analysis are indicated with grey dotted lines.}
    \label{fig:wavelength_bins}
\end{figure}

A single bin spanning the entire wavelength range is used to define the white-light light-curve. We use 24 wavelength bins of the stellar spectra to define spectroscopic light-curves. We attempt to use constant 20 nm width bins, with minor modifications in width to avoid bin edges being located near absorption features in the stellar spectrum. We use three narrower bins of $10$ nm ($9$ nm for the central bin) width to explore the core and wings of the Na I doublet feature at 589 nm. We choose to use a single 30 nm bin to cover the entire teluric \ch{O2} A band feature at 762 nm, to avoid any bin edges falling in the wings of the feature. This bin happens to also include the K I doublet at 770 nm, which is found in the wing of the \ch{O2} absorption. We present example stellar spectra from the first night annotated with the adopted wavelength bins in Fig. \ref{fig:wavelength_bins}.

We create normalized differential white-light and spectroscopic light-curves by summing the flux within the relevant wavelength bins for the target, dividing by the corresponding flux of the comparison star, and normalizing by the median out-of-transit flux.

\section{Data Analysis}
\label{sec:fitting}

\subsection{Transit model}

We fit the differential light-curves using analytic light curves \citep{mandel2002analytic} from the {\tt Batman} package \citep{kreidberg2015batman}, with detrending based on physical basis functions to remove red noise (see Sect.\,\ref{sec:detrending}).
We use the Limb-Darkening Toolkit ({\tt LDTK}) package \citep{parviainen2015ldtk} implemented through {\tt Tiberius} to compute the limb-darkening coefficients using {\tt PHOENIX} models \citep{husser2013new}, based on the following stellar parameters of HAT-P-44 from \citet{hartman2014hat}: $T_{\mathrm{eff}}=5295\pm100$ K, $\log{g}=4.46\pm0.06$, and Fe/H$=0.33\pm0.10$ . 

First we fit the white-light light curves for the planet-to-star radius ratio $R_p/R_*$, the system inclination $i$, the mid-transit time $T_c$, the ratio of semi-major axis to stellar radius $a/R_*$, and the first quadratic limb-darkening coefficient $u_1$, using only wide uniform priors. We fix the period as ($P = 4.30119043$ d) based on the most precise literature value from \citet{kokori2023exoclock}. 
\new{We perform fits using different limb-darkening parametrizations, including linear and quadratic limb-darkening laws with each parameter fixed or fitted (using the Kipping parametrization \citep{kipping2013efficient} to avoid degeneracy when both $u_1$ and $u_2$ are fitted).
We let the data inform us on the most appropriate level of detail by choosing the fit with the best Bayesian Information Criterion (BIC), which slightly favoured quadratic limb-darkening with $u_2$ fixed and $u_1$ fitted for the white-light light-curves in both nights, and significantly favoured a quadratic law with both $u_2$ and $u_1$ fixed for the spectroscopic light-curves over any other method.}

We present the results of these white-light light curve fits in Table \ref{tab:planet_parameters}. All the parameters are consistent between the nights, except for $u_1$, which is strongly discrepant between the nights. The value from the first night is consistent with the calculated value of $u_1=0.584\pm0.003$. We attribute the discrepancy to the scattered light feature we see during the second night in Fig.\,\ref{fig:extraction-frames} and Fig.\,\ref{fig:ancillary_plots2}. 
For the spectroscopic light-curves we fix $i$, $T_c$, and $a/R_*$ to the best-fitting values from the white-light light-curve 
\new{of the first night data}.

\begin{table*}
    \centering
    \begin{tabular}{c|c|c|c|c}
       \hline
        Parameter & Prior & Night 1 Fitted Value & Night 2 Fitted Value &  \citet{hartman2014hat}\\
        \hline
        $R_p/R_*$ & Uniform & $0.1367\substack{+0.0010 \\ -0.0008}$\, & $0.1363\substack{+0.0011 \\ -0.0010}$\, & $0.1343 \pm 0.0010$\\
        $a/R_*$ & Uniform & $11.95\substack{+0.13 \\ -0.22}$\, & $11.91\substack{+0.20 \\ -0.21}$\,  & $11.50\substack{+0.50 \\ -0.80}$\,\\
        $i$ $(^\circ)$ & Uniform & $89.2\substack{+0.5 \\ -0.4}$\, & $88.9\substack{+0.5 \\ -0.4}$\, & $89.1\pm0.4$\, \\
        $u_1$ & Uniform & $0.58\pm0.02$ & $0.45\pm0.02$\\
        $u_2$ & Fixed & 0.0042 & 0.0042\\
        \hline
    \end{tabular} 
    \caption{Parameter values from white-light light-curve fits for both nights. Values for the planet-to-star radius ratio $R_p/R_*$, the ratio of semi-major axis to stellar radius $a/R_*$, and the system inclination $i$, are consistent between nights, while the values for the first quadratic limb-darkening coefficient $u_1$ are inconsistent, an effect we attribute to the scattering event during the transit in the second night. The calculated value for $u_1$ is $0.584\pm0.003$, consistent with the first night.}
    \label{tab:planet_parameters}
\end{table*}

\subsection{Light-curve fitting}
\label{sec:detrending}
For both nights, we fit the white-light and spectroscopic light-curves with the transit model and \new{a range of physical basis functions}. We investigate all combinations of linear and quadratic basis functions including sky background, full width at half-maximum (FWHM), and x and y pixel shifts, \new{plus linear in time and airmass. We compare the resulting BIC to determine the best combination of detrending parameters. We choose the detrending parameters independently for the white-light and spectroscopic light-curves, letting the BIC guide our choice in both cases.}

\begin{figure}
    \centering
    \includegraphics[width=0.49\textwidth]{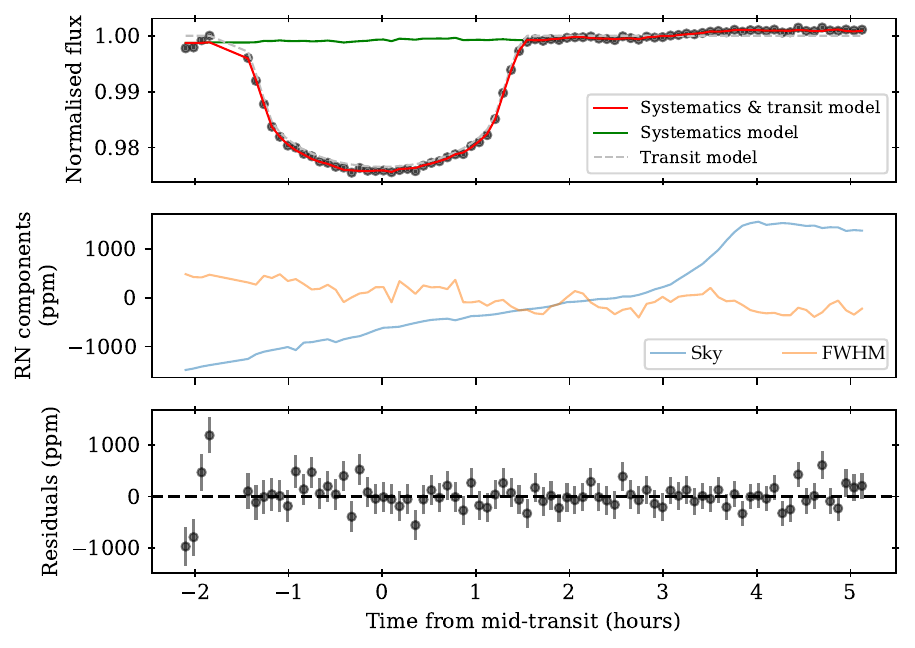}
    \caption{Top panel: White-light light-curve of our first night of observation of HAT-P-44b (black data points), with an analytic transit model (grey dotted), polynomial systematics model (green), and the combined fit to the data (red). Middle panel: Detrending components contributing to the systematics model: quadratic in sky background (poly 1, blue) and \new{quadratic} in FWHM (poly 2, orange). Bottom panel: Residuals from the combined systematics and transit model.}
    \label{fig:WL_night1}
\end{figure}
\begin{figure*}
    \centering
    \includegraphics[width=0.97\textwidth]{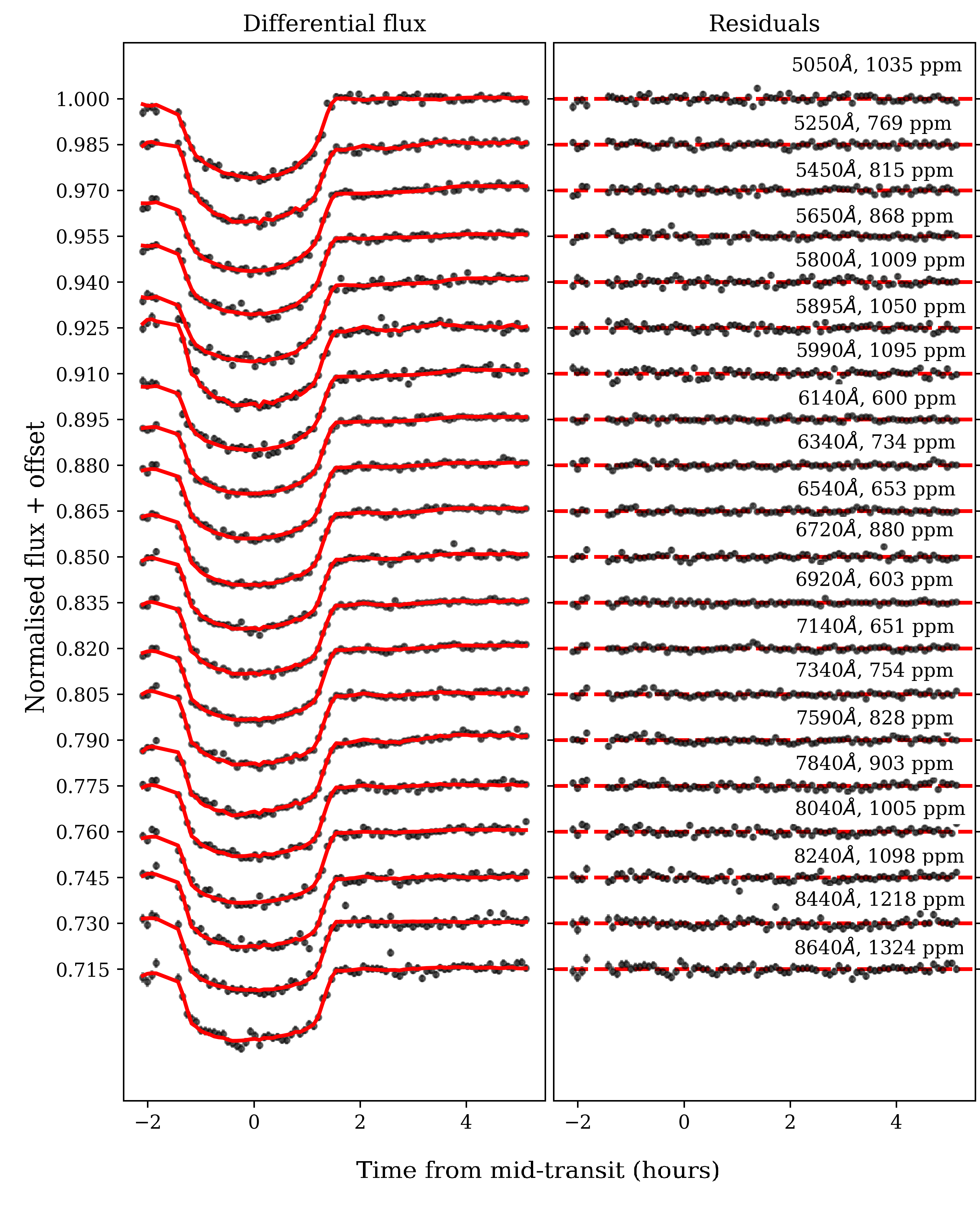}
    \caption{Spectroscopic light-curve fits from our first night of observation of HAT-P-44b (black data points) with the combined transit model and polynomial detrending fit (red) with different $y$ offsets. Residuals to these fits are displayed in the right-hand panel, with the central wavelength and residual rms displayed \new{above} each fit.}
    \label{fig:wb_night1}
\end{figure*}

\new{In the first night we find that the detrending parameters resulting in the best BIC \newerer{for the white-light light curve} are quadratic sky and quadratic FWHM.
These parameters are able to account for the variable seeing and remove the kink in the out-of-transit baseline corresponding to moonset, which we correct for using quadratic sky background detrending. We clip the fifth and final four data points from the light-curves, as they are strongly affected by clouds, visible as dips in the normalized flux in Fig. \ref{fig:ancillary_plots}. For the spectroscopic light-curves we find that a simpler detrending with linear sky and linear FWHM gives the best BIC, by accounting for the impact of variable seeing.}
We present our white-light light-curve fit in Fig. \ref{fig:WL_night1} and our spectroscopic fits in Fig.\,\ref{fig:wb_night1}. \newest{In order to assess the impact of red noise, we include plots in Fig.\,\ref{fig:rms_vs_bins} demonstrating how the root mean square (RMS) residuals of the spectroscopic light-curve fits vary as a function of binning.  For all but one bin we find that the residuals are consistent with $1/\sqrt{N}$ behaviour of white noise. The only bin to deviate from this behaviour is the bin centred at 759 nm, which contains the telluric \ch{O2} feature. This bin does not impact our interpretation of the transmission spectrum.}


\begin{figure}
    \centering
    \includegraphics[width=0.49\textwidth]{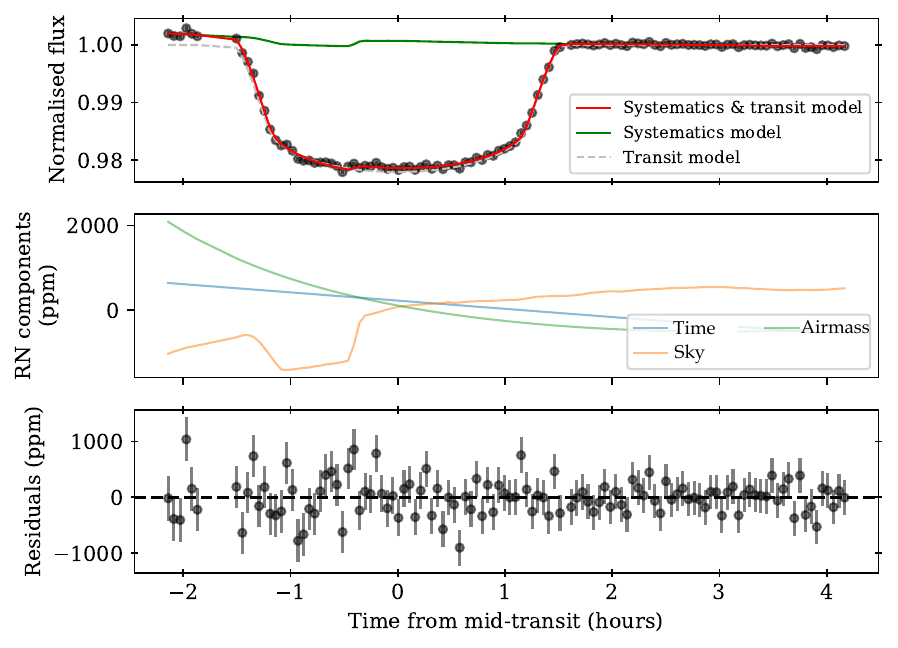}
    \caption{Top panel: White-light light-curve of our first night of observation of HAT-P-44b, with an analytic transit model (grey dotted), polynomial systematics model (green), and the combined fit to the data (red). Middle panel: Detrending components contributing to the systematics model: linear in time (poly 1, blue), \new{linear} in sky background (poly 2, orange), and linear in airmass (poly 3, green). Bottom panel: Residuals from the combined systematics and transit model.}
    \label{fig:WL_night2}
\end{figure}

\begin{figure*}
    \centering
    \includegraphics[width=0.97\textwidth]{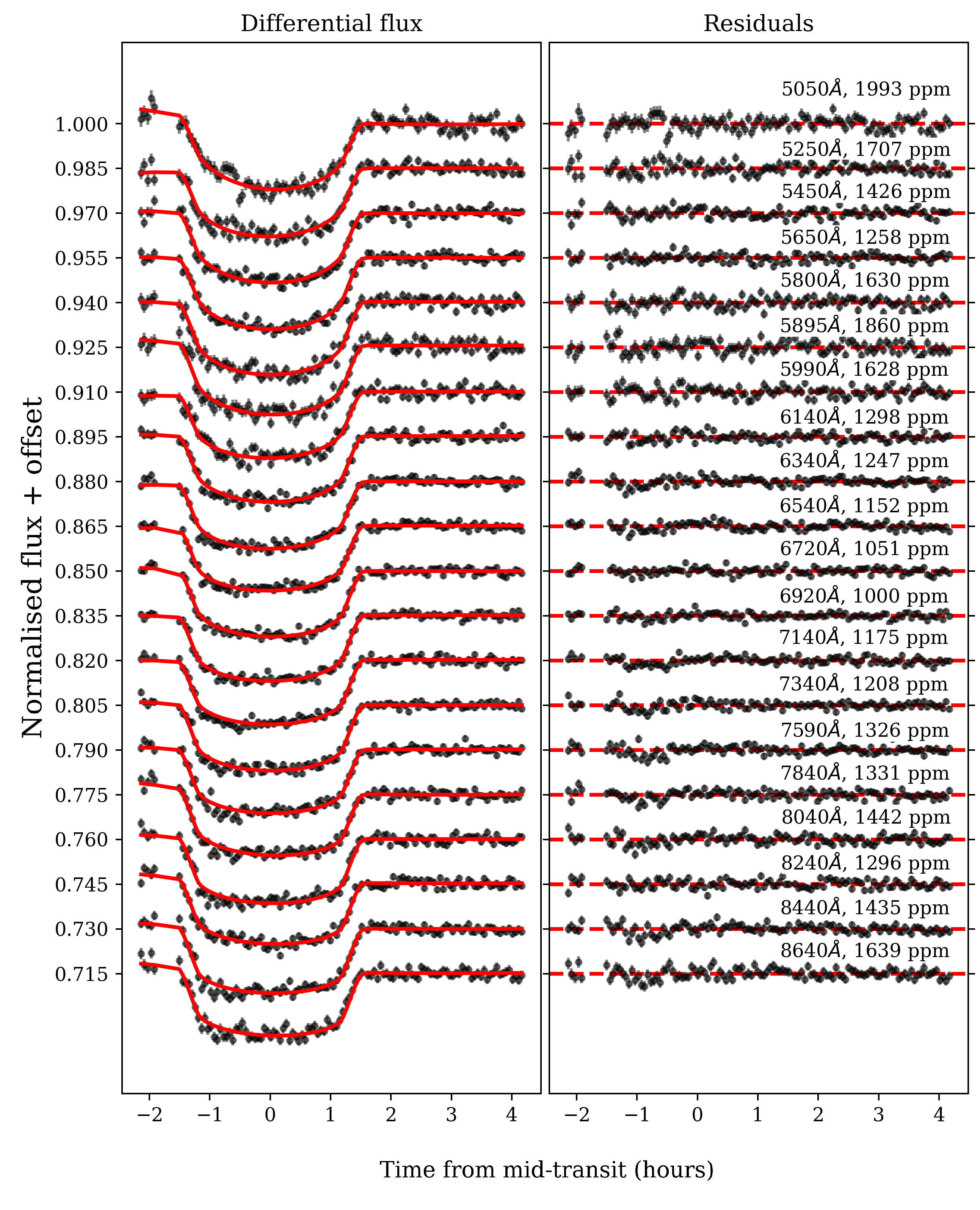}
    \caption{Spectroscopic light-curve fits from our second night of observation of HAT-P-44b (black data points) with the combined transit model and polynomial detrending fit (red) with different $y$ offsets. Residuals to these fits are displayed in the right-hand panel, with the central wavelength and residual rms displayed \new{above} each fit.}
    \label{fig:wb_night2}
\end{figure*}

In the second night, we have an interval of spatially-localised
background enhancement \new{on the detector} during the transit, which is clearly visible in Fig. \ref{fig:extraction-frames} and Fig. \ref{fig:ancillary_plots2}. 
\new{With careful background subtraction we can eliminate most of its impact on the differential white-light light-curve, but we still see a small impact on the residuals. However, we see a much stronger impact on the spectroscopic light-curves, given the complicated 2D structure of the scattering feature (see Fig. \ref{fig:extraction-frames}), differentially affecting the blue wavelengths. Since the scattered light feature only effects one of the stars, we lose the power of differential spectro-photometry at these times. Using best BIC detrending, we find that linear basis functions of time, airmass, and sky background are favoured for the white-light light-curve, while the combination of just linear airmass and linear sky is favoured for the spectroscopic light-curves.}
We also clip the final hour of observation, which is affected by moonset and poor seeing. \new{We also attempt fits where the drop in flux due to clouds for five exposures just before transit are both included and cropped, finding the inclusion has a negligible effect on the resulting transmission spectrum.} The results of our white-light and spectroscopic light-curve fits are illustrated in Fig. \ref{fig:WL_night2} and Fig. \ref{fig:wb_night2} respectively.

We find generally poorer fits 
in the lower signal-to-noise bins at the edge of the detector, so we restrict our transmission spectrum to a wavelength range of 495--874 nm.


 The fit for each light-curve was performed using  Markov Chain Monte Carlo (MCMC) from the {\tt emcee} package \citep{foreman2013emcee}, with an initial run with 5000 steps and 12 walkers per parameter, followed by a production run with a further 5000 steps. \new{10000 steps were used for both the initial run and production run for fitting the white-light light-curves.} A Nelder-Mead algorithm \citep{nelder1965simple} was employed for initial parameter optimization, after which photometric uncertainties were rescaled.

\begin{table}
    \centering
    \begin{tabular}{c|c|c}
       \hline
        Night 1 Spectroscopic Fit & $k$  & $\delta$\,BIC   \\
        \hline
        Linear Sky, Linear FWHM  & 4 & +0 \\
        Linear Sky, Linear FWHM, Linear Airmass  & 5 & +70 \\
        Linear Sky, Linear Airmass  & 4 & +76 \\
        Quadratic Sky, Linear FWHM & 5 & +87 \\
        Linear Sky  & 3 & +166\\
        Linear Time  & 3 & +189\\
        Quadratic Sky, Quadratic FWHM  & 6 & +205\\
        \hline 
        Quadratic LD, Both Fixed  & 4 & +0\\
        Quadratic LD, $u_2$ Fitted  & 5 & +158\\
        Quadratic LD, Both Fitted  & 6 & +295\\
        Linear LD, Fixed  & 4 & +237\\
        Linear LD, Fitted  & 5 & +163\\
        \hline 
    \end{tabular}
    \caption{\newer{The combined BICs of different systematic models and limb-darkening parametrizations on the spectroscopic light-curve fits to the first night, relative to the best BIC model ($\delta$ BIC = 0), with the number of free parameters ($k$).}}
    \label{tab:BICs}
\end{table}

\subsection{Transmission spectrum}

\begin{figure}
    \centering
    \includegraphics[width=0.49\textwidth]{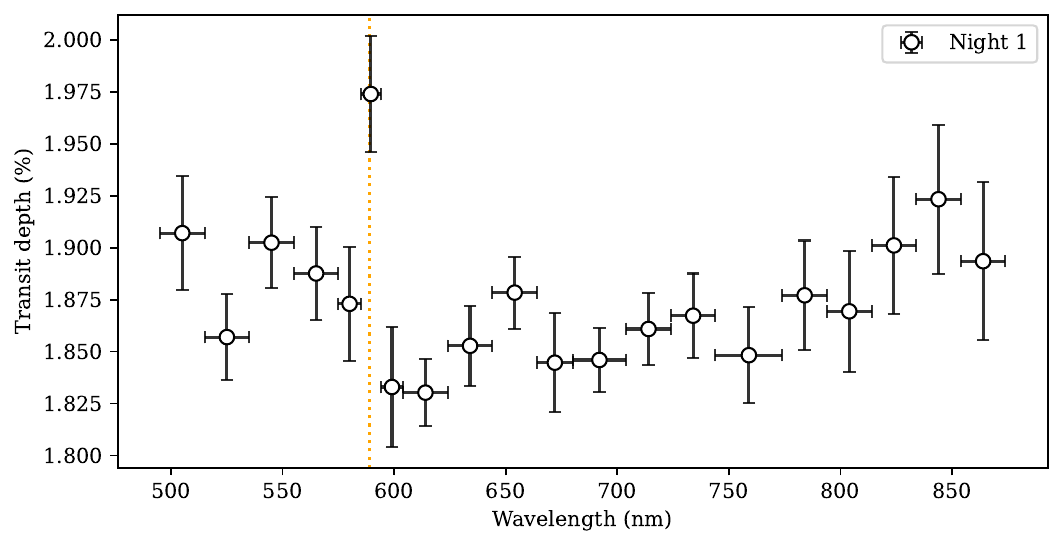}
    \includegraphics[width=0.49\textwidth]{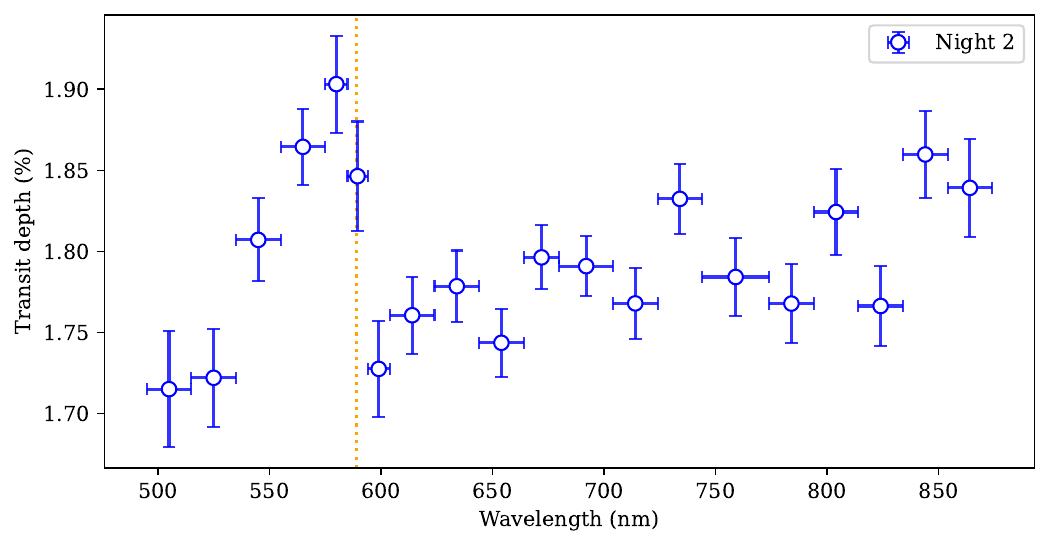}
    \includegraphics[width=0.49\textwidth]{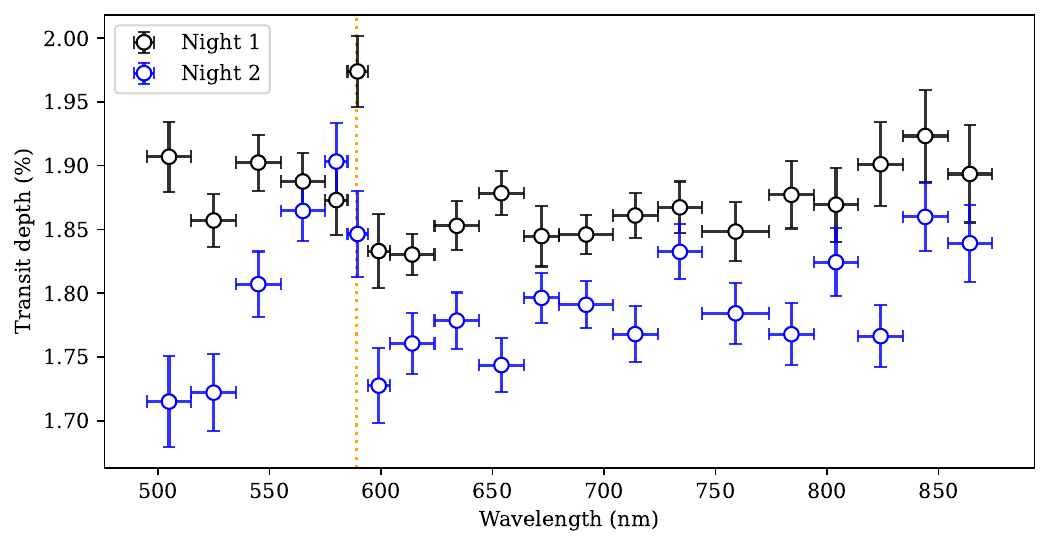}
    \caption{Transmission spectrum of HAT-P-44b from spectroscopic light-curve fits from the first night (black) and the second night (blue), using the refit limb darkening coefficients. The wavelength of Na I absorption at 589 nm is highlighted with a yellow dotted line.}
    \label{fig:spectrum_both_nights}
\end{figure}

\begin{table}
    \centering
    \begin{tabular}{c|c|c}
       \hline
        Wavelength & Night 1  & Night 2  \\
        (nm) &  $R_p/R_*$ &  $R_p/R_*$ \\
        \hline
        495--515  & $0.1381\substack{+0.0010 \\ -0.0010}$\, & $0.1310\substack{+0.0014 \\ -0.0014}$\, \\
        515--535  & $0.1363\substack{+0.0008 \\ -0.0008}$\, & $0.1312\substack{+0.0011 \\ -0.0012}$\,\\
        535--555  & $0.1379\substack{+0.0008 \\ -0.0008}$\, & $0.1344\substack{+0.0009 \\ -0.0010}$\,\\
        555--575  & $0.1373\substack{+0.0008 \\ -0.0008}$\, & $0.1365\substack{+0.0009 \\ -0.0009}$\,\\
        575--585  & $0.1369\substack{+0.0010 \\ -0.0010}$\, & $0.1380\substack{+0.0011 \\ -0.0011}$\,\\
        585--594  & $0.1405\substack{+0.0010 \\ -0.0010}$\, & $0.1359\substack{+0.0012 \\ -0.0012}$\,\\
        594--604  & $0.1354\substack{+0.0011 \\ -0.0011}$\, & $0.1314\substack{+0.0011 \\ -0.0011}$\,\\
        604--624  & $0.1353\substack{+0.0006 \\ -0.0006}$\, & $0.1327\substack{+0.0009 \\ -0.0009}$\,\\
        624--644  & $0.1361\substack{+0.0007 \\ -0.0007}$\, & $0.1334\substack{+0.0008 \\ -0.0008}$\,\\
        644--664  & $0.1371\substack{+0.0006 \\ -0.0006}$\, & $0.1320\substack{+0.0008 \\ -0.0008}$\,\\
        664--680  & $0.1358\substack{+0.0009 \\ -0.0009}$\, & $0.1340\substack{+0.0007 \\ -0.0007}$\,\\
        680--704  & $0.1359\substack{+0.0006 \\ -0.0006}$\, & $0.1338\substack{+0.0007 \\ -0.0007}$\,\\
        704--724  & $0.1364\substack{+0.0006 \\ -0.0006}$\, & $0.1330\substack{+0.0008 \\ -0.0008}$\,\\
        724--744  & $0.1366\substack{+0.0007 \\ -0.0007}$\, & $0.1354\substack{+0.0008 \\ -0.0008}$\,\\
        744--774  & $0.1360\substack{+0.0008 \\ -0.0009}$\, & $0.1336\substack{+0.0009 \\ -0.0009}$\,\\
        774--794  & $0.1370\substack{+0.0010 \\ -0.0010}$\, & $0.1330\substack{+0.0009 \\ -0.0009}$\,\\
        794--814  & $0.1367\substack{+0.0011 \\ -0.0011}$\, & $0.1351\substack{+0.0010 \\ -0.0010}$\,\\
        814--834  & $0.1379\substack{+0.0012 \\ -0.0012}$\, & $0.1329\substack{+0.0009 \\ -0.0009}$\,\\
        834--854  & $0.1387\substack{+0.0013 \\ -0.0013}$\, & $0.1364\substack{+0.0010 \\ -0.0010}$\,\\
        854--874  & $0.1376\substack{+0.0014 \\ -0.0014}$\, & $0.1356\substack{+0.0011 \\ -0.0011}$\,\\
        \hline
    \end{tabular}
    \caption{Tabulated transmission spectra from both nights of observation, plotted in Figure \ref{fig:spectrum_both_nights}.}
    \label{tab:transmission_spectrum}
\end{table}

We construct our highlighted transmission spectra for both nights using the transit depth of the spectroscopic light-curves, presented in Table \ref{tab:transmission_spectrum} and Figure \ref{fig:spectrum_both_nights}. Our recovered spectra cover a wavelength range of 495--874 nm using $\sim$20 nm bins,  
\new{with evidence for absorption at the wavelength of the sodium doublet in night 1. When comparing the spectra, we see quite a strong disagreement, with night 2 displaying broad absorption at blue wavelengths which cannot readily be attributed to any atmospheric feature. Given the strong scattering feature that impacted a large amount of the in-transit spectrum (Fig.\,\ref{fig:extraction-frames}), we believe that despite our best efforts in background subtraction and detrending, we were not able to remove its impact on the transmission spectrum. We therefore repeated the fits, cropping all exposures impacted by the scattering event. This results in cropping out almost half the transit including the entire ingress. Using the best BIC detrending for this cropped light-curve does not result in an improved transmission spectrum. We therefore conclude that the need to either detrend out or crop the scattering event results in a loss of information in the spectroscopic light-curves making it impossible to recover an accurate transmission spectrum for this night of observation. We therefore proceed using only our transmission spectrum from the first night.} 

We achieve a mean transit depth precision of \new{246 ppm in the first night, which is less than the} 288 ppm expected transit signal of one scale height of the planetary atmosphere, \new{permitting detailed atmospheric characterization.}


\subsection{Testing the sodium detection}
\label{sec:sodium}

\begin{figure}
    \centering
    \includegraphics[width=0.49\textwidth]{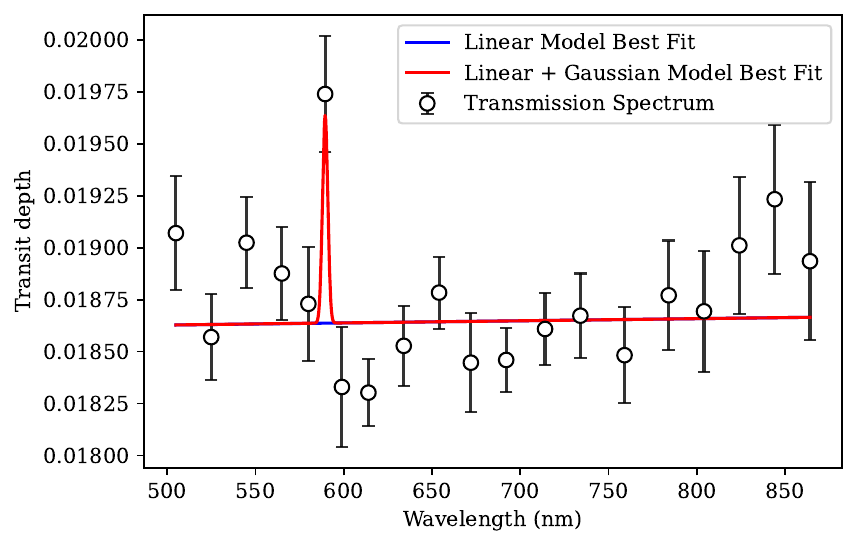}
    \caption{Transmission spectrum of HAT-P-44b from the first night with simple parametric model fits, including a linear model (blue) with the sodium bin masked, and a linear model plus a Gaussian feature with fixed mean centred on the sodium feature and fitted width (red) fitted to the whole spectrum.}
    \label{fig:sodium-slope-test}
\end{figure}

\new{With a strong feature at the wavelength of Na I absorption in our spectrum from the first night, we conduct simple tests to estimate the significance and reliability of the detection. Following \citet{nikolov2016vlt} and \citet{alderson2020lrg}, we fit a linear polynomial to the transmission spectrum with the sodium bin masked and obtain a 3.9 $\sigma$ significance based on the difference between the fit and the sodium bin (see Fig.\,\ref{fig:sodium-slope-test}). We also fit the entire spectrum with \newer{three simple models: a linear polynomial; a linear polynomial with a Gaussian feature fixed at 589.3 nm with fixed 6nm width and fitted amplitude; and a Voigt feature fixed at 589.3 nm with free width, amplitude, and Lorentzian half-width at half-maximum. The Voigt feature represents giving the sodium freedom to be pressure-broadened, while the Gaussian feature represents a narrow feature with the width set by the spectral resolution.
The linear plus fixed Gaussian sodium feature gives the best reduced $\chi_\nu^2$ of 1.33, representing a 3.5 $\sigma$ improvement on the linear fit.} We believe the lack of improvement of the Voigt fit to be representative of an inability to directly detect broadening as a result of the broad bins set by the high FWHM of our observations. \newest{Switching from a linear polynomial to a simple Rayleigh scattering slope \citep{pont2008detection} does not improve the fit because it cannot reproduce the overall redward shape of the spectrum (with a positive temperature).} 
We plot a selection of \newest{our parametric} fits in Fig. \ref{fig:sodium-slope-test}.}

\begin{figure}
    \centering
    \includegraphics[width=0.49\textwidth]{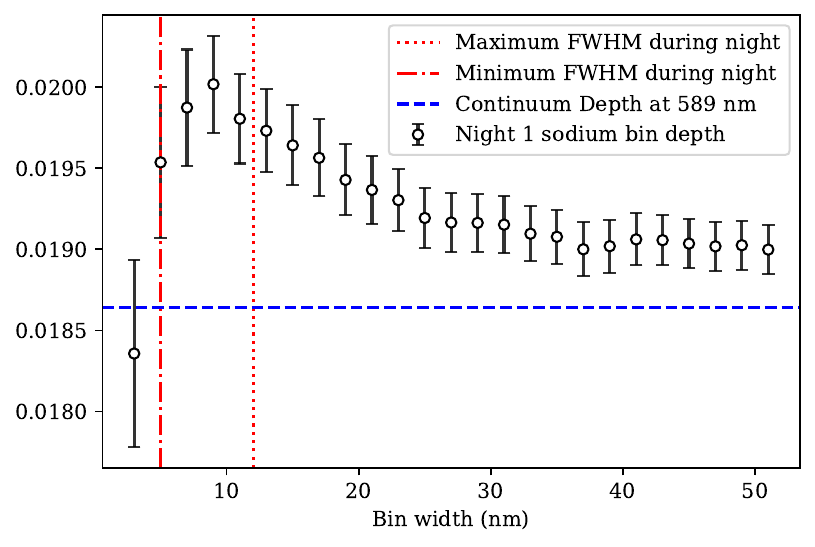}
    \caption{Transit depth of a binned light-curve centred on the sodium absorption feature at 589 nm as a function of bin width in the first night. These results are consistent with a feature with the width of the spectral resolution (which varies from 5 to 11 nm throughout the night), with decreasing transit depth at increasing bin width above this resolution. In the second night, this trend is much less clear, likely caused by the equally high transit depth in an adjacent bin.}
    \label{fig:bin-width-test}
\end{figure}

\new{We conducted a further investigation into the presence of sodium by fitting light-curves with various bin widths centred on the Na I feature at 589 nm. This can be used to identify the width of the feature and test that it is centred at the correct wavelength. For a narrow feature, the expected behaviour is a sharp decrease in transit depth with increasing bin width, approaching the surrounding continuum level \citep[see][]{alderson2020lrg}.}
\new{This is the behaviour we observe in the first night, with a more gradual decrease in width than in \citeauthor{alderson2020lrg}, remaining above the continuum level at 50 nm bin width, likely due to the strength of the feature we observe. The spectral resolution deteriorates to $\sim 12$ nm initially due to the poor seeing, just above the 9 nm bin width corresponding to the highest transit depth. As can be seen in Figure \ref{fig:bin-width-test}, the transit depth decreases with bin width above this, as expected.
The effect of the stellar sodium feature being smoothed out into adjacent bins due to the strongly varying spectral resolution throughout the night is particularly apparent in the narrowest bins. The FWHM detrending does a good job of eliminating this effect in wider bins, but struggles in the smallest bin.}


We repeated the analysis on the K I feature at 768 nm, but see no evidence for increased absorption. The close proximity of the strong telluric \ch{O2} feature can result in large systematic errors when using narrow bins, particularly with variable FWHM. When using a 20 nm bin, as is used for the rest of the spectrum, the transit depth is consistent with the adjacent bins (see Fig. \ref{fig:spectrum_both_nights}).

\section{Atmospheric Retrieval}
\label{sec:retrieval}

\subsection{Retrieval setup}

\begin{figure*}
    \centering
    \includegraphics[width=0.97\textwidth]{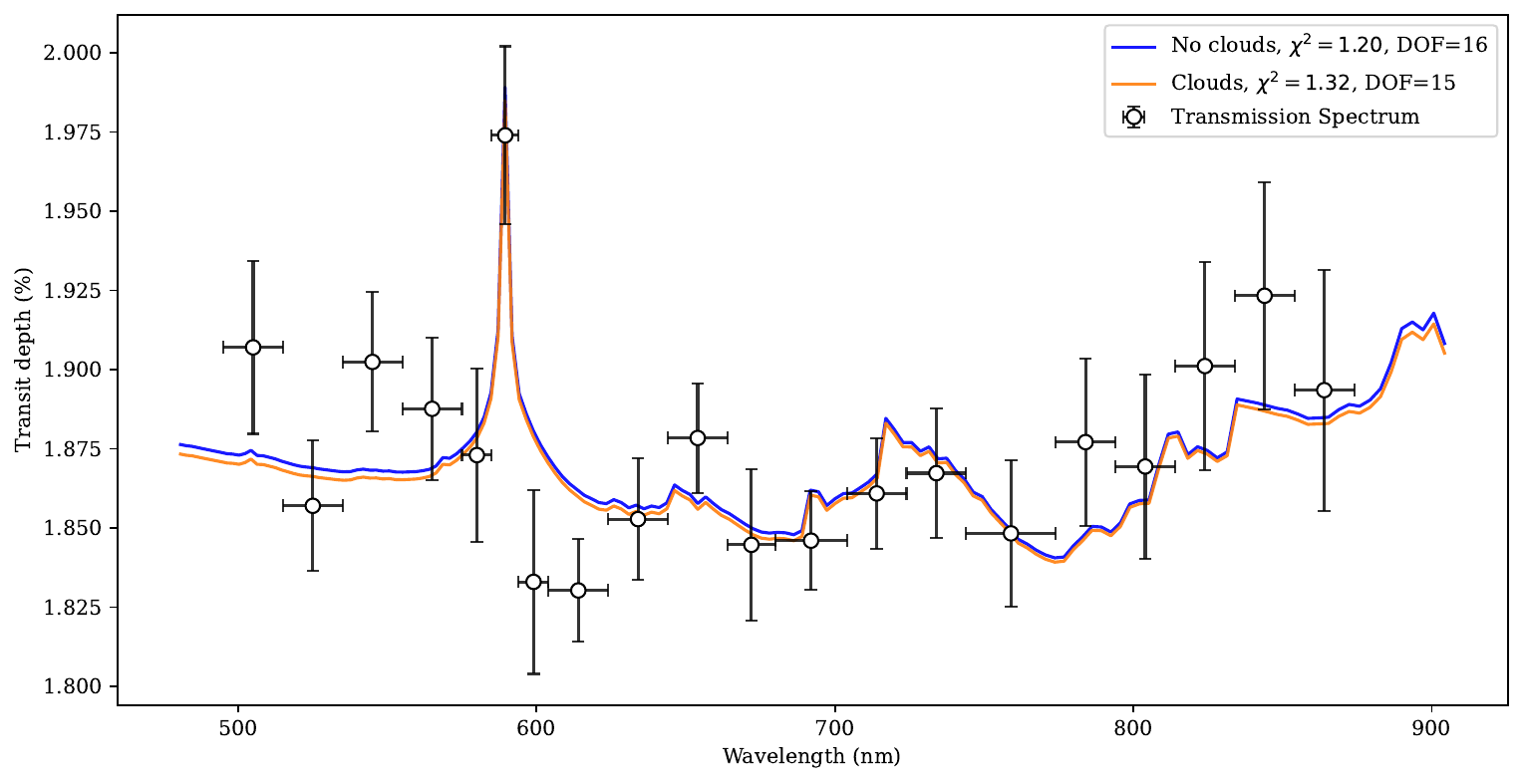}
    \caption{Transmission spectrum from the first night with (black data points) and the best-fitting {\tt petitRADTRANS} model from the highest evidence atmospheric retrieval (blue). The spectrum demonstrates absorption due to sodium (the sharp winged feature at 589 nm), Rayleigh scattering without haze-enhancement (the downwards slope at the blue end of the spectrum), and \ch{H2O} and \ch{CH4} absorption (the upwards slope at the red end of the spectrum, and the small feature at 730 nm). We also plot the best-fitting model from the second highest evidence atmospheric retrieval (orange), which also includes a grey cloud layer as a free parameter, with the $2 \sigma$ minimum cloud-top pressure at $\sim$ 1 mbar.}
    \label{fig:spectrum_best_fit}
\end{figure*}

We carry out an atmosphere retrieval analysis using the {\tt petitRADTRANS 2.7.7} package \citep{molliere2019petitradtrans,Nasedkin2024} to interpret our transmission spectrum from the first night. 
{\tt petitRADTRANS} uses Bayesian nested sampling with {\tt MultiNest} \citep{feroz2009multinest} and radiative transfer with correlated-$k$ opacity tables to compute and fit synthetic transmission spectra to the observed spectrum.

\new{We perform a collection of retrievals with different aerosol parameterizations and including various opacity sources. In all of our models} we use an isothermal atmosphere and equilibrium chemistry, with a fixed C/O ratio of 0.55, a Gaussian prior on the gravity with mean and standard deviation from \citet{hartman2014hat}, and uniform priors on the planetary radius, isotherm temperature, and metallicity. We include combinations of opacity from \ch{H2O} \citep{polyansky2018exomol}, \ch{CH4} \citep{yurchenko2017hybrid}, and Na \citep{piskunov1995vald}, as well as collisionally induced absorption from \ch{H2-H2} and \ch{H2-He} and Rayleigh scattering from \ch{H2} and He. We parameterize the chemical abundances using metallicity, which is scaled based on the solar abundances of \citet{asplund2009chemical}. We fix the C/O ratio, \new{since the wavelength range covered by our spectrum is insufficiently sensitive to relative abundances of carbon- and oxygen-bearing species.}

\new{For our aerosol parametrizations, we \newer{run retrievals including} opacity from haze, modelled by the expected Rayleigh scattering for \ch{H2} and \ch{He} multiplied by a fitted haze factor, and grey cloud, modelled using an opaque grey cloud deck at a fitted cloud-top pressure, as well as retrievals including neither source of opacity as well as both. By including both sources,} we are effectively assuming that aerosol opacity in this case is driven by some combination of condensate clouds with large particle sizes fixed below some altitude, as is observed in spectral muting of absorption features in transmission spectra, plus some photochemically produced aerosols which are sufficiently small to be well modelled by Rayleigh scattering. 


In previous optical transmission spectra, we have observed scattering slopes even steeper than Rayleigh \citep{sing2011gran,alderson2020lrg,luque2020obliquity,ahrer2022lrg}, often attributed to photochemical hazes \citep{kawashima2019theoretical,ohno2020super}. Therefore we also run a retrieval in which we permit an arbitrary scattering slope of the form $\lambda = S^{-\alpha}$, with the gradient permitted to vary between 4 and 20 (it is fixed to 4 for Rayleigh scattering). Both the gradient $\alpha$ and the multiplicative factor $S$ are fitted parameters in this retrieval.

\subsection{Retrieval results}

\begin{figure*}
    \centering
    \includegraphics[width=0.9\textwidth]{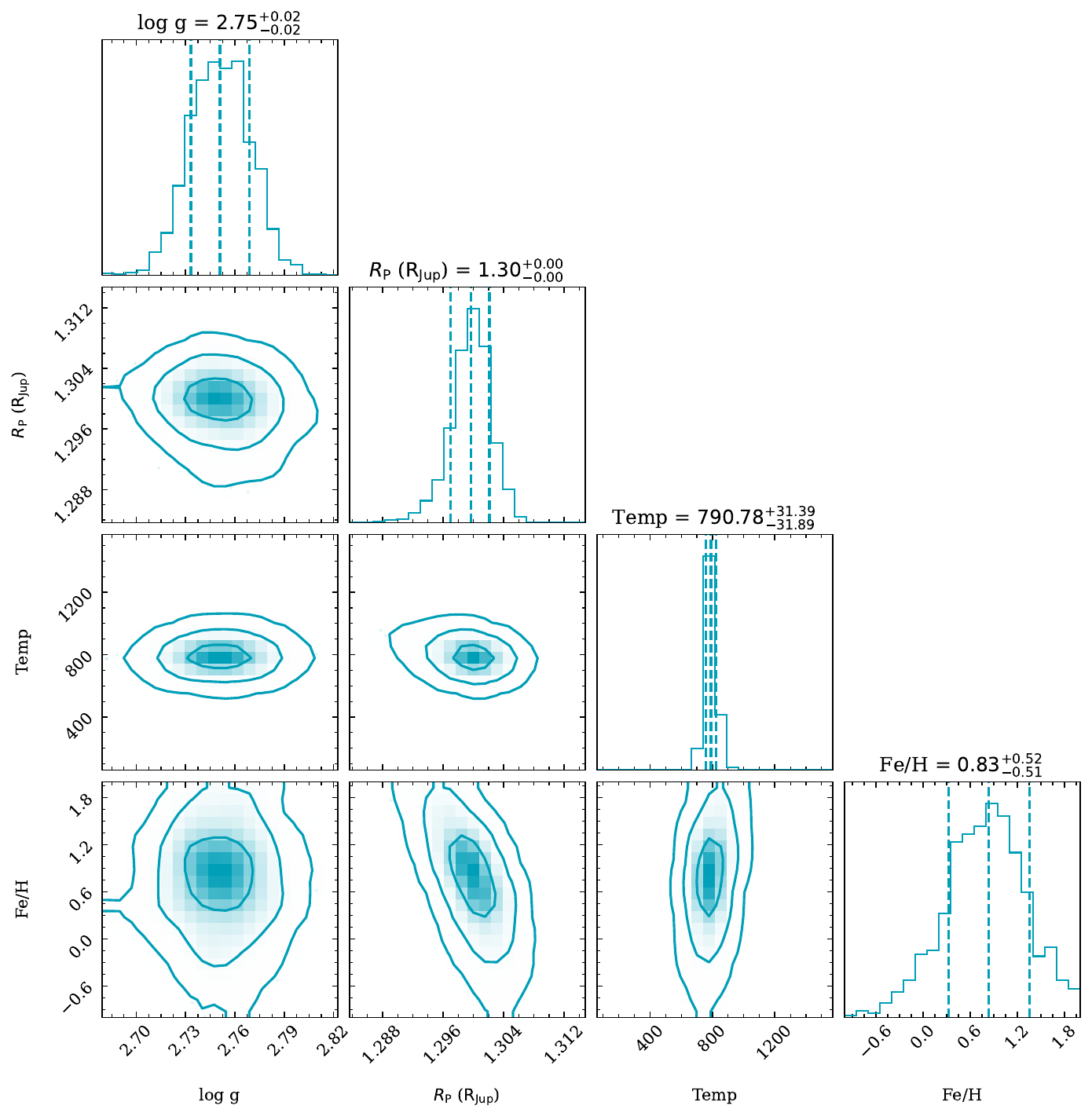}
    \caption{\new{Corner plot of posterior distributions of retrieval parameter values obtained by fitting the transmission spectrum from the first night with the best evidence {\tt petitRADTRANS} atmospheric model using a Rayleigh scattering slope, equilibrium chemistry, and an isothermal atmosphere, as depicted in blue in Fig. \ref{fig:spectrum_best_fit}.}}
    \label{fig:posteriors_rayleigh}
\end{figure*}

\new{Our retrieval with the highest Bayesian evidence includes absorption from Na, \ch{H2O}, and \ch{CH4}, and no aerosol opacity from cloud or haze (see Table\,\ref{tab:lnZs}).} This very simple setup gives us a \newer{concordant spectrum using opacity from Na, \ch{H2O}, and \ch{CH4} with a single atmospheric metallicity} of $7${\raisebox{0.5ex}{\tiny$\substack{+16 \\ -5}$}}$\times$ solar, and gives tight constraints of the temperature of $790\pm30$ K. This is cooler than the 1100 K equilibrium temperature, as is commonly observed in 1D isothermal retrievals \citep{MacDonald2020,Welbanks2022}. We plot the best fitting model from this retrieval in Fig. \ref{fig:spectrum_best_fit} and the posteriors in Fig.\,\ref{fig:posteriors_rayleigh}. \newest{We also include the full set of fits to the transmission spectrum in Fig.\,\ref{fig:spectrum_all_fits}.}


\begin{table*}
    \centering
    \begin{tabular}{c|c|c|c|c|c}
       \hline
        Retrieval / Simple Model & $\ln{Z}$  & $B_m$ & $\chi_\nu^2$ & DOF & Rejection $\sigma$\\
        \hline
        No clouds & -20.1 & - & 1.20 & 16 & \\
        Clouds & -20.1 & 1.0 & 1.32 & 15 & -\\
        Clouds + Haze & -20.4 & 1.6 & 1.45 & 14 & -\\
        Clouds + super-Rayleigh & -23.2 & 25.1 & 1.68 & 13 & 2.07\\
        \hline 
        Linear Fit & - & - & 2.01 & 18 & 3.53\\
        Linear + Gaussian Fit & - & - & 1.33 & 17 & 1.52\\
        Linear + Voigt Fit & - & - & 1.50 & 15 & -\\
        \hline 
    \end{tabular}
    \caption{\newer{The Bayesian evidence ($\ln{Z}$), Bayes factor ($B_m$), reduced $\chi_\nu^2$, degrees of freedom (DOF), and significance of being rejected relative to the no clouds retrieval reference model ($\sigma$ reject), of the different retrievals and simple model fits applied to the transmission spectrum. For retrievals, the significance of rejection is converted from the Bayes factor via the p-value using $p=\frac{1}{1+B_m}$, while a likelihood ratio test using the $\chi_\nu^2$ values is used for the simple model fits. The rejection $\sigma$ is not given for retrievals where the Bayes factor is too small for it to be meaningful.}}
    \label{tab:lnZs}
\end{table*}

\begin{figure*}
    \centering
    \includegraphics[width=0.97\textwidth]{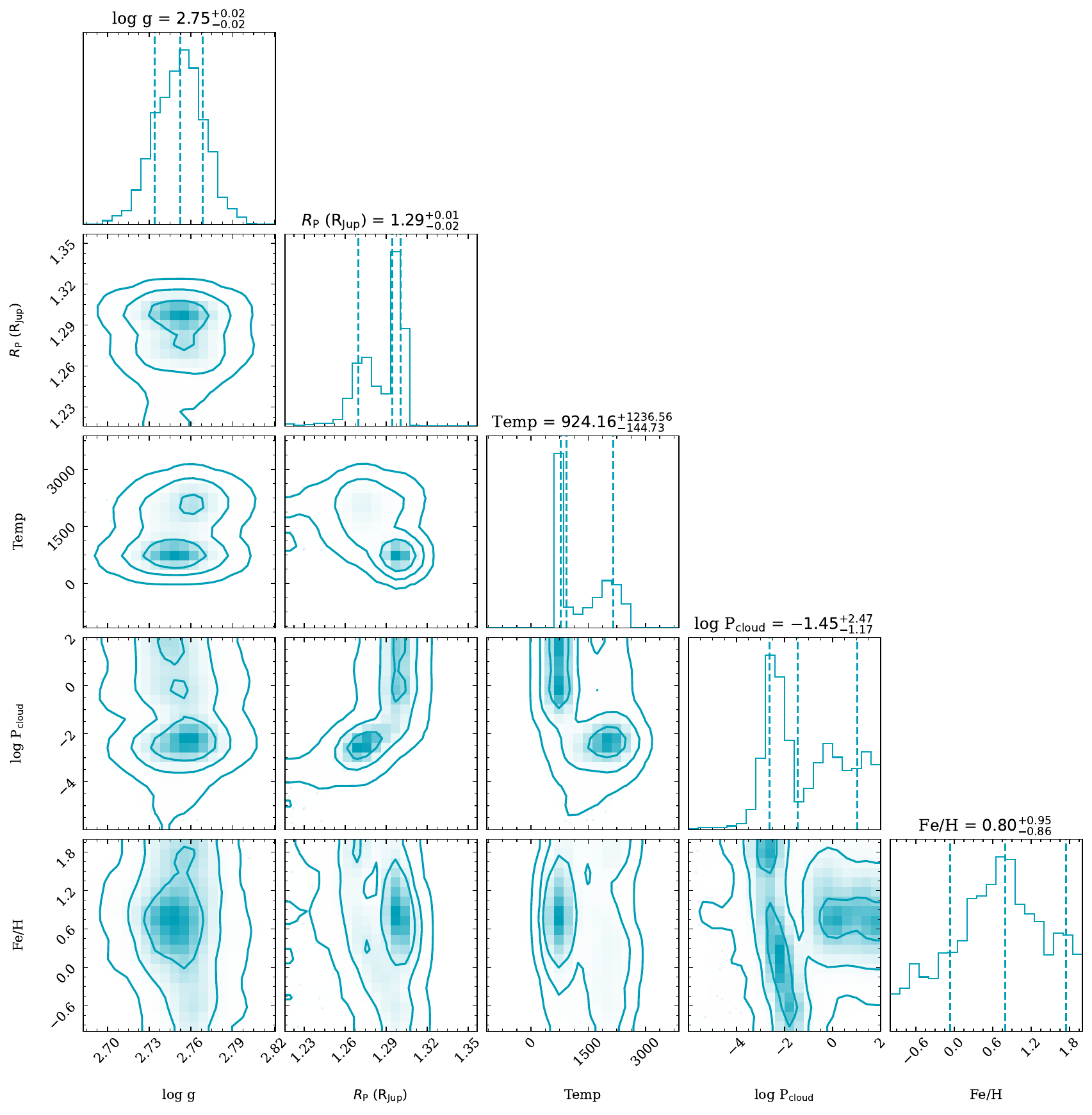}
    \caption{\new{Corner plot of posterior distributions of retrieval parameter values obtained by fitting the transmission spectrum from the first night with the second best evidence {\tt petitRADTRANS} atmospheric model, including an isothermal atmosphere, equilibrium chemistry, a Rayleigh scattering slope, and a grey cloud deck, as depicted in orange in Fig. \ref{fig:spectrum_best_fit}.}}
    \label{fig:posteriors_free}
\end{figure*}

\newer{We see nearly identical Bayesian evidence when we include opacity from a grey cloud deck, which yields an identical best-fitting model with 
essentially no
muting from clouds (see Fig. \ref{fig:spectrum_best_fit}). This places a $2 \sigma$ upper limit on the cloud-top pressure of $\sim$ 1 mbar, ruling out high altitude clouds such as those responsible for muting the optical transmission spectrum of HATS-46b \citep{ahrer2023lrg}. The near equal evidence of this retrieval demonstrates that we are agnostic to the inclusion of mid-altitude ($>1$ mbar) cloud. However, the posterior distributions in Fig.\,\ref{fig:posteriors_free} highlight that a degeneracy is present in the temperature, metallicity, and cloud-top pressure posteriors when clouds are considered.} \newer{There are two modes to this retrieval, with the best-fitting mode being the same as the retrieval without clouds, and an additional mode including muted water features due to clouds at $\sim 10$ mbar and unconstrained metallicity and a higher temperature of $\sim1600$ K.}

\newer{Comparing the evidence with the other retrieval setups in Table\,\ref{tab:lnZs} shows a 
slight
preference for 
not including
a Rayleigh haze enhancement, and notably disfavouring  a free (super-Rayleigh) haze slope (with a Bayes factor of 25). 
We present the posteriors of the cloud-free retrieval in Fig. \ref{fig:posteriors_rayleigh}, and the cloudy retrieval in Fig. \ref{fig:posteriors_free}. We summarise our Bayesian evidences, best-fit reduced $\chi_\nu^2$ values, degrees of freedom, and detection significances of all our retrievals and simple model fits in Table \ref{tab:lnZs}.} \newest{Repeating the atmospheric retrieval while omitting the data point centred at 759 nm containing the telluric \ch{O2} feature resulted in no change to the posteriors or best fit spectrum, highlighting that any red noise associated with this feature does not impact our interpretation of the transmission spectrum.}

\section{Discussion}
\label{sec:discussion}

\subsection{The detection of sodium}
Our 
analysis using simple parametric model fits results in a \new{3.9$\sigma$ detection of Na I absorption in HAT-P-44b, using our sub-scale height precision transmission spectrum from the first night of observation} (see Sec. \ref{sec:sodium} for details). The observed sodium feature spans $\>5$ scale heights of transit depth even with a broad 9 nm bin. 


Previous ground-based optical transmission spectra have given diverse results for sodium absorption, including non-detection of sodium \citep[e.g.,][]{kirk2017rayleigh,ahrer2023lrg}, detection of the narrow core of a sodium feature \citep[e.g.,][]{redfield2008sodium,chen2018gtc,alderson2020lrg}, and detection of the pressure-broadened wings of the sodium feature \citep[][]{nikolov2016vlt,nikolov2018absolute}. This is typically attributed to varying cloud altitude suppressing the feature strength and obscuring the feature wings or core with a scattering slope. Detection of a strong sodium feature is usually taken as evidence for a clear atmosphere, especially when obvious pressure-broadened wings are also detected. 

The low and variable resolution of the first night (5--11 nm) hinders direct detection of pressure-broadening in HAT-P-44b, with no improvement in $\chi_\nu^2$ from using a linear plus Voigt fit over a linear plus fixed-width Gaussian fit for the whole spectrum. However, the very strong transit depth even with a relatively broad 9 nm bin requires a very strong and hence probably broad sodium feature. We emphasize our 9 nm bin is much broader than the typical 3 nm bin sodium detections \citep[e.g.][]{alderson2020lrg}. As highlighted in our retrieval analysis in Sec. \ref{sec:retrieval}, our observed features \new{imply a haze-free atmosphere, with additional absorption from some combination of \ch{H2O} and \ch{CH4}, \newer{and optionally mid-altitude clouds (2 $\sigma$ upper limit of 1 mbar on the cloud altitude)}, to accurately fit the shape of the spectrum. The presence of a strong sodium feature and the detectability of the weak molecular features at these wavelengths strongly fit the narrative of a cloud-free atmosphere.}

With the detection of a strong sodium feature, HAT-P-44b joins a small group of exoplanets with potentially cloud-free upper atmospheres: WASP-39b \citep{fischer2016hst,nikolov2016vlt,sing2016continuum}, WASP-96b \citep{nikolov2018absolute,mcgruder2022access}, WASP-62b \citep{alam2021evidence}, and WASP-94 Ab \citep{ahrer2022lrg}. These planets have proven ideal for high SNR infra-red follow-up with JWST \citep{ahrer2023early,alderson2023early,feinstein2023early,radica2023awesome,rustamkulov2023early,taylor2023awesome}, which can provide exceptional chemical and structural detail.

Despite our detection of sodium, we do not detect potassium in HAT-P-44b, although we stress that due to the proximity of potassium to the telluric \ch{O2} feature, our spectrum is poorly sensitive to potassium absorption. We see evidence for potassium depletion in the transmission spectra of WASP-21b \citep{alderson2020lrg}, WASP-96b \citep{nikolov2018absolute,taylor2023awesome} and WASP-62b \citep{alam2021evidence}. In contrast we have strong detections of potassium in WASP-39b \citep{fischer2016hst,feinstein2023early}. The cause of this discrepancy is unknown, the depletion of potassium could be due to some primordial depletion, or sequestering of potassium by condensates like KCl. 

\subsection{Aerosols and evidence for methane and water}


\new{Outside of the sodium feature, we see some structure in the observed spectrum, including both increasing transit depth towards both the blue and red ends of the spectrum (see Fig. \ref{fig:spectrum_best_fit}). Blueward slopes in the optical spectra of hot Jupiter atmospheres are common}, and can be evidence for scattering slopes, or stellar in-homogeneity arising from activity. The gradient of the slope combined with the strength of the sodium feature can explain the aerosol properties of the atmosphere. A slope gradient consistent with Rayleigh scattering is typical of small particles, whether that is the \ch{H2} gas, haze, or mineral condensate particles \citep[e.g.][]{des2008rayleigh}. A flat spectrum and suppressed sodium absorption indicates large particles in the form of mineral clouds \citep[e.g.][]{ackerman2001precipitating,fortney2005effect}. A steeper slope can arise from a photochemical haze \citep{kawashima2019theoretical,ohno2020super}, a combination of cloud and haze layers \citep{pont2013prevalence}, sulfide clouds \citep{pinhas2017signatures}, or stellar activity such as unocculted spots or occulted faculae \citep{mccullough2014water,oshagh2014impact,espinoza2019access}. \new{Redward slopes are typically the result of molecular absorption, like \ch{H2O}, \ch{CH4}, TiO, and VO.}

\newer{We use atmospheric retrievals to investigate these features in HAT-P-44b, 
allowing us to investigate 
opacity from Rayleigh scattering (with or without haze enhancement), a grey-cloud deck, and 
molecules.
The best-fit model from our highest evidence retrieval gives an improvement in goodness of fit equivalent to being favoured at 1.5 $\sigma$ compared to our parametric fit ($\chi_\nu^2$ of 19.2 with 16 degrees of freedom for the retrieval versus $\chi_\nu^2$ of 22.6 with 17 degrees of freedom for the linear plus Gaussian fit).}

\new{In this retrieval, our bluewards slope is attributed to Rayleigh scattering from \ch{H2} and He gas, while the increased absorption at the red end is attributed to molecular absorption from \ch{H2O} and \ch{CH4}, with abundances consistent with the sodium abundance in chemical equilibrium. Combined, these features point towards a cool (800 K), aerosol-free, super-Solar metallicity ($7${\raisebox{0.5ex}{\tiny$\substack{+16 \\ -5}$}}$\times$Solar) upper atmosphere. This remains the favoured explanation even when a grey cloud deck is permitted, although adding a grey cloud deck into the model adds degeneracies between metallicity, cloud-top pressure, temperature, and radius, complicating the picture and preventing a direct measurement of metallicity or temperature.} The addition of a Rayleigh opacity enhancement term, representative of haze particles, is very slightly disfavoured in the Bayesian evidence, while inclusion of a free (i.e. potentially super-Rayleigh) scattering slope parameterization is moderately disfavoured.

\begin{figure*}
    \centering
    \includegraphics[width=0.97\textwidth]{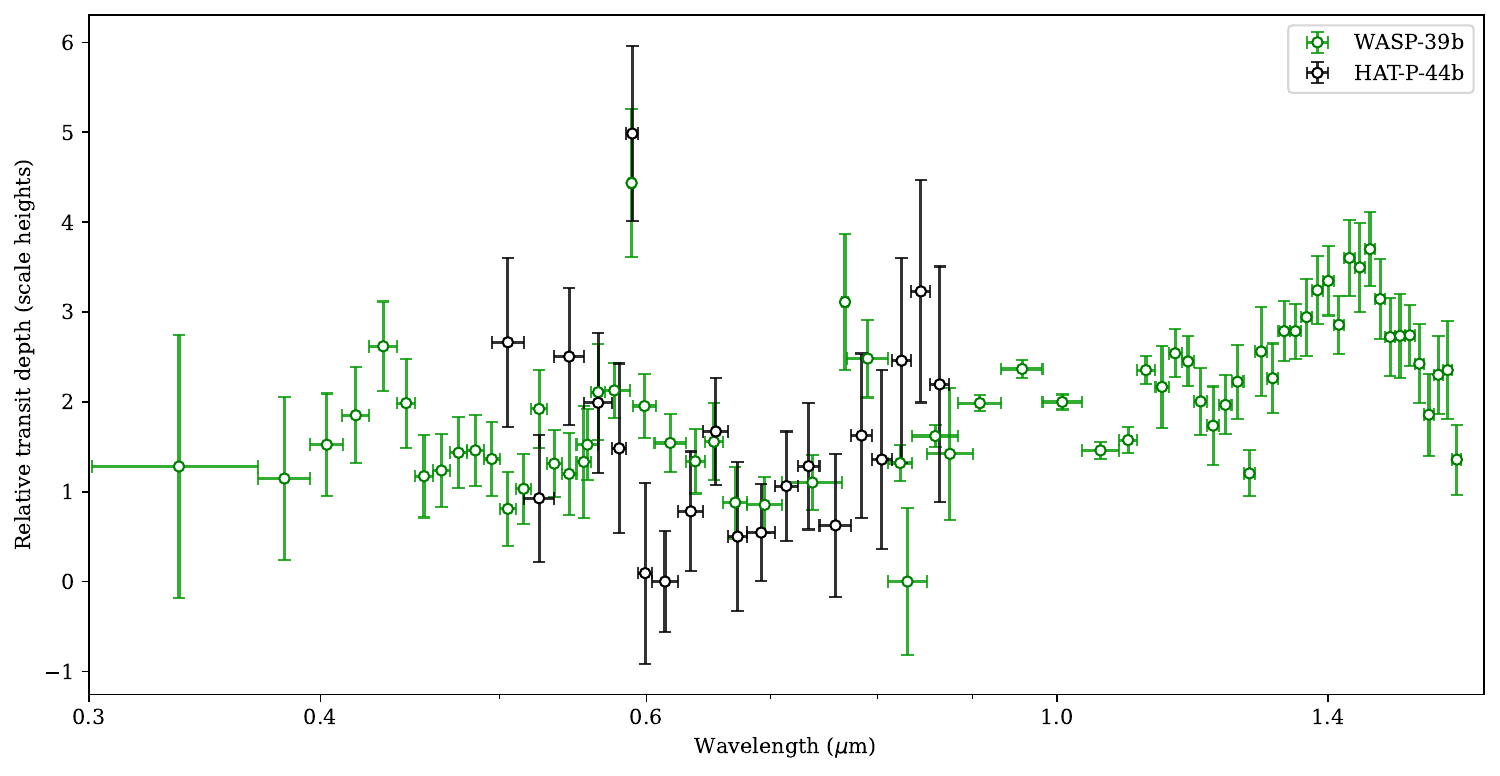}
    \caption{Transmission spectrum of HAT-P-44b compared against the transmission spectrum of WASP-39b, a similar inflated hot Saturn. The WASP-39b spectrum includes VLT, HST, and WHT data from \citet{fischer2016hst,nikolov2016vlt,sing2016continuum,wakeford2017complete} as combined by \citet{kirk2019lrg}. The spectra have similar sodium absorption and continuum levels.}
    \label{fig:w39_hp44_compairons}
\end{figure*}

Stellar activity can also cause bluewards slopes in optical transmission spectra that mimic Rayleigh scattering slopes \citep{mccullough2014water,oshagh2014impact}. This is notable in planets around more active stars like HD 189733b, where the super-Rayleigh slope can also be explained with occulted faculae or unocculted spots. Stellar activity is a potential alternative explanation for the slope in HAT-P-18b \citep[][]{fournier2024near},
but HAT-P-44 is not an active star, 
with a median $\log R_{\mathrm{HK}} = -5.247 \pm 0.058 \pm 0.10$ \citep{hartman2014hat} \new{and a low $v\sin{i}=0.2\pm0.5$ $\mathrm{kms^{-1}}$}, indicating low levels of chromospheric activity. \newer{We note however that it is possible for calcium emission to be masked by a calcium outflow from some planets \citep{haswell2020dispersed}, and a low $v\sin{i}$ can potentially arise from low stellar inclination.}
%


\newer{Our observation of features in the transmission spectrum of HAT-P-44b appears consistent with the results of the reconnaissance spectrum obtained by \citet{oelkers2025ground}, which suggests HAT-P-44b may be clear based on the variance in its transit depth in different wavelength bins.}

\subsection{HAT-P-44b in context}

\begin{table*}
\begin{tabular}{lllllll}
       \hline
                 & HAT-P-44b & WASP-39b      & WASP-21b       & WASP-96b      & WASP-62b      & WASP-94Ab     \\
                 \hline
$T_\mathrm{{eq}}$ (K)          & $1110${\raisebox{0.5ex}{\tiny$\substack{+50 \\ -30}$}}   & $1116\pm14$    & $1340\pm30$   & $1285\pm40$          & $1475${\raisebox{0.5ex}{\tiny$\substack{+25 \\ -20}$}}          & $1604${\raisebox{0.5ex}{\tiny$\substack{+25 \\ -22}$}}           \\
Gravity ($\mathrm{ms^{-2}}$)       & $5.7\pm1.1$       & $4.3\pm0.6$           & $5.6\pm0.3$         & $8.3\pm0.1$           &    $7.4\pm1.4$           & $3.8\pm0.4$            \\
Stellar $T_\mathrm{{eff}}$ (K) & 5300      & 5400          & 5800           & 5500          & 6200          & 6200           \\
Scattering slope & Rayleigh  & Rayleigh      & Super-Rayleigh & Rayleigh      & Not observed  & Super-Rayleigh \\
Haziness         & Weak  & Weak          & Moderate       & Weak          & Weak          & Moderate       \\
Sodium           & Strong    & Strong, broad & Strong, narrow         & Strong, broad & Strong, broad & Strong, broad \\
Potassium        & None$^*$     & Present       & None$^*$          & Tentative         & None          & Unknown        \\
       \hline\\
    
\end{tabular}
\\ $^*$ From only ground-based data, so may be impacted by the telluric \ch{O2} feature.
    \caption{Comparison of hot Jupiters with evidence for relatively cloud-free atmospheres from their optical transmission spectra due to strong sodium absorption, scattering slopes, or both. We use moderate haziness to refer to planets with strong (relative to the sodium core, or obscuring the sodium wings) or super-Rayliegh scattering slopes, and weak haziness to planets with clear sodium wings and weak or non-detected scattering slopes. Strong haziness would refer to planets with strong Rayleigh or super-Rayleigh scattering slopes that lead to no or weak sodium features like HAT-P-18b \citep{kirk2017rayleigh}. We summarise the results of the following atmospheric characterizations: HAT-P-44b (this paper), WASP-39b \citep{fischer2016hst,nikolov2016vlt,sing2016continuum,kirk2019lrg,feinstein2023early}, WASP-21b \citep{alderson2020lrg,chen2020detection}, WASP-96b \citep{nikolov2018absolute,mcgruder2022access,radica2023awesome,taylor2023awesome}, WASP-62b \citep{alam2021evidence}, and WASP-94Ab \citep{ahrer2022lrg,ahrer2024atmospheric}. The discovery papers for the included exoplanets are as follows: HAT-P-44b \citep{hartman2014hat}, WASP-39b \citep{faedi2011wasp}, WASP-21b \citep{bouchy2010wasp}, WASP-96b \citep{hellier2014transiting}, WASP-62b \citep{hellier2012seven}, WASP-94Ab \citep{neveu2014wasp}.}
\label{tab:clear_planet_comparison}

\end{table*}

\new{With the combination of a strong sodium feature and hints of \ch{H2} gas Rayleigh scattering and \ch{H2O} and \ch{CH4} absorption, we infer a relatively haze- and cloud-free upper atmosphere, placing a 2$\sigma$ upper limit on a grey cloud-top pressure at 1 mbar, and find the scattering slope strength to be consistent with 800\,K \ch{H2} gas scattering. We therefore conclude that the parts of the atmosphere we probe using transmission spectroscopy are likely lacking in small particle aerosols, as well as large particle aerosols above 1 mbar.}

In this regard, the optical transmission spectrum of HAT-P-44b has similarities to those of WASP-21b \citep{alderson2020lrg,chen2020detection}, WASP-39b \citep{fischer2016hst,nikolov2016vlt,sing2016continuum,kirk2019lrg}, WASP-94 Ab \citep{ahrer2022lrg,ahrer2024atmospheric}, WASP-96b \citep{nikolov2018absolute,mcgruder2022access}, and WASP-62b \citep{alam2021evidence}. A comparison of these planets' planetary parameters and the features of their optical transmission spectra are presented in Tab. \ref{tab:clear_planet_comparison}.

WASP-39b and WASP-21b are notable points of comparison for HAT-P-44b, with similar equilibrium temperatures and surface gravities, with WASP-39b also possessing a similar host star to HAT-P-44. In terms of its optical spectrum, HAT-P-44b sits closer to WASP-39b in aerosol properties, with a stronger sodium feature and weaker slope than WASP-21b indicative of less haze. We depict the spectra of both HAT-P-44b and WASP-39b in Fig. \ref{fig:w39_hp44_compairons} for comparison. Like WASP-21b, HAT-P-44b shows no evidence of potassium in its transmission spectrum, although both data sets are ground-based and are potentially affected by the same issue with the telluric \ch{O2} feature. WASP-39b on the other hand does have potassium absorption detected using space-based observatons \citep{fischer2016hst,sing2016continuum,feinstein2023early}. 

\citet{gao2020aerosol} predict that silicates are the dominant aerosol species responsible for muting spectral features at these temperatures, while photochemical hazes are implicated in super-Rayleigh slopes \citep{kawashima2019theoretical,ohno2020super}. The similarities between the optical spectra of the sibling planets HAT-P-44b and WASP-39b are therefore suggestive that these processes are driven mainly by these planetary properties: gravity, temperature, and stellar host. 
On the other hand, HATS-46b ($T_\mathrm{{eq}}=1100$\,K, $g=5.1$ $\mathrm{ms^{-2}}$, stellar $T_\mathrm{{eff}} = 5400$\,K) is another close sibling to both WASP-39b and HAT-P-44b yet has a cloudy optical transmission spectrum \citep{ahrer2023lrg}.
Looking at the hotter, higher gravity side of hot Saturns, we see a featureless spectrum indicative of clouds for WASP-49b \citep{lendl2016fors2}, which is otherwise a close analogue ($T_\mathrm{{eq}}=1400$\,K, $g=7.8$ $\mathrm{ms^{-2}}$, stellar $T_\mathrm{{eff}} = 5600$\,K) to WASP-96b in these parameters.

The impact of both temperature and gravity on cloudiness has been explored with both computer modelling and statistical analysis of observations, with generally higher temperature and gravity associated with less cloud \citep{heng2016cloudiness,stevenson2016quantifying,fu2017statistical,dymont2022cleaning,estrela2022temperature}. The growing roster of optical transmission spectra make it apparent that while there is some island of relatively cloud-free hot Saturns, with temperatures 1000--1500 K and surface gravities 4--8 $\mathrm{ms^{-2}}$, whether a given exoplanet exists in this island cannot be predicted from these properties alone. This could be explained by the chemical inventories of those planets, or complicated dependencies on various planetary and stellar parameters involving photochemistry, 3D dynamics, and aerosol microphysics \citep[see][and references within]{gao2021aerosols}.

\section{Conclusions}
\label{sec:conclusion}

We present the low-resolution optical transmission spectrum of \newer{the inflated hot Saturn HAT-P-44b, the} tenth planet in the LRG-BEASTS survey. Making use of the ACAM instrument on the WHT and \new{using data from one of two nights of observation (the other being compromised by systematics from scattered light)}, we obtain a transmission spectrum from 495--874 nm with an average transit depth precision of \new{246} ppm using $\sim$ 20 nm bins, which is 0.85$\times$ the signal strength of one scale height. Using a comparison star and 
detrending based on physical parameters including sky background, airmass, and FWHM, we are able to effectively remove systematic effects from variable seeing, scattered light, moonset, and atmospheric extinction, with the median RMS-noise equal to the expected photon noise.

We detect strong sodium absorption spanning $\sim 5$ scale heights in the spectrum of HAT-P-44b with 3.9$\sigma$ confidence using a broad 9 nm bin. \new{We also see evidence for Rayleigh scattering from \ch{H2} gas and absorption due to \ch{H2O} and \ch{CH4} in our highest evidence retrieval, consistent with a relatively cloud-free, 800 K atmosphere with a super-solar metallicity ($7${\raisebox{0.5ex}{\tiny$\substack{+16 \\ -5}$}}$\times$ solar). A super-Rayleigh slope indicative of strong photochemical haze like those observed in WASP-21b and WASP-94Ab is disfavoured at 2.1 $\sigma$ significance, while the 2 $\sigma$ upper limit cloud-top pressure is placed at $\sim 1$ mbar.} 


In this regard, the optical transmission spectrum of HAT-P-44b resembles that of WASP-39b, a close sibling in surface gravity, equilibrium temperature, and stellar host. While it is compelling to suggest that this similarity in planetary properties is the cause of these planets' similar aerosol properties, it is worth noting that another close sibling in temperature, gravity, and stellar host, HATS-46b, shows evidence for clouds in the transmission spectrum. 
The physics and chemistry determining aerosol properties is therefore evidently more complex than a simple function of these three planetary parameters, despite HAT-P-44b apparently joining a group of similar 1000--1500\,K hot Saturns with relatively cloud-free atmospheres.

\new{The presence of identifiable features in the} atmosphere of HAT-P-44b makes it an ideal candidate for further atmospheric characterization, particularly high SNR infra-red follow-up with JWST, as has been successfully done for other relatively cloud-free hot Saturns WASP-39b and WASP-96b. Extending the transmission spectrum further into the blue would also permit more detailed analysis of the scattering slope and any potential impact of stellar activity.




\section*{Acknowledgements}

The William Herschel Telescope is operated on the island of La Palma by the Isaac Newton Group 
of Telescopes in the Spanish Observatorio del Roque de los Muchachos of the Instituto de Astrofísica de Canarias. 

This research has made use of the NASA Exoplanet Archive, which is operated by the California Institute of Technology, under contract with the National Aeronautics and Space Administration under the Exoplanet Exploration Program. ABC acknowledges studentship support from the UK Science and Technology Facilities Council (STFC), and PJW acknowledges support from STFC under consolidated grants ST/P000495/1, ST/T000406/1 and ST/X001121/1. JK acknowledges financial support from Imperial College London through an Imperial College Research Fellowship grant.

\section*{Data Availability}

Tabulated transmission spectra, reduced light-curves, fitting statistics, and input files for both nights are available on Zenodo at  https://doi.org/10.5281/zenodo.18744185.




\bibliographystyle{mnras}
\bibliography{main} 




\appendix

\section{Tests of Background Subtraction}

\begin{figure*}
    \centering
    \includegraphics[width=0.85\textwidth]{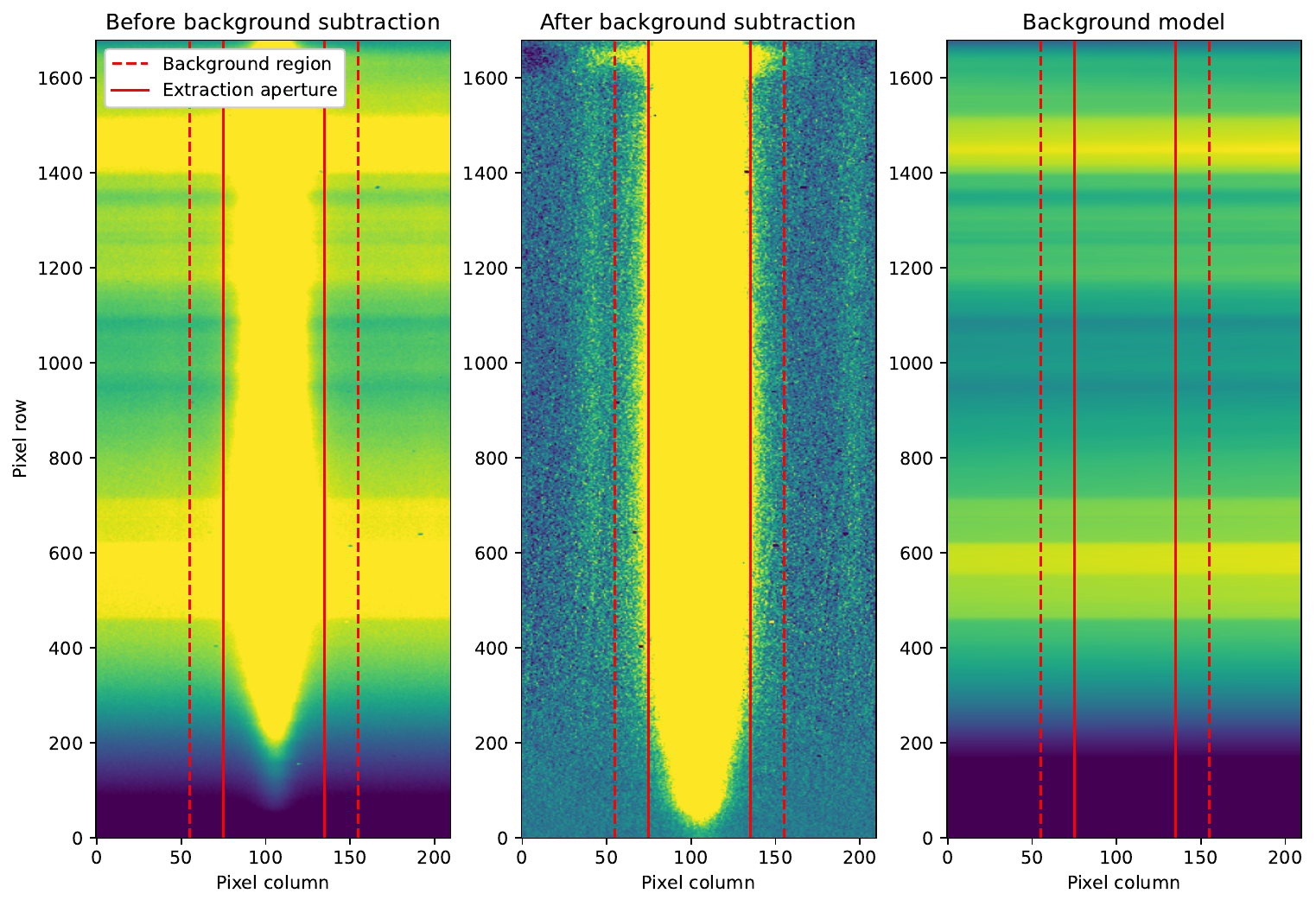}
    \includegraphics[width=0.85\textwidth]{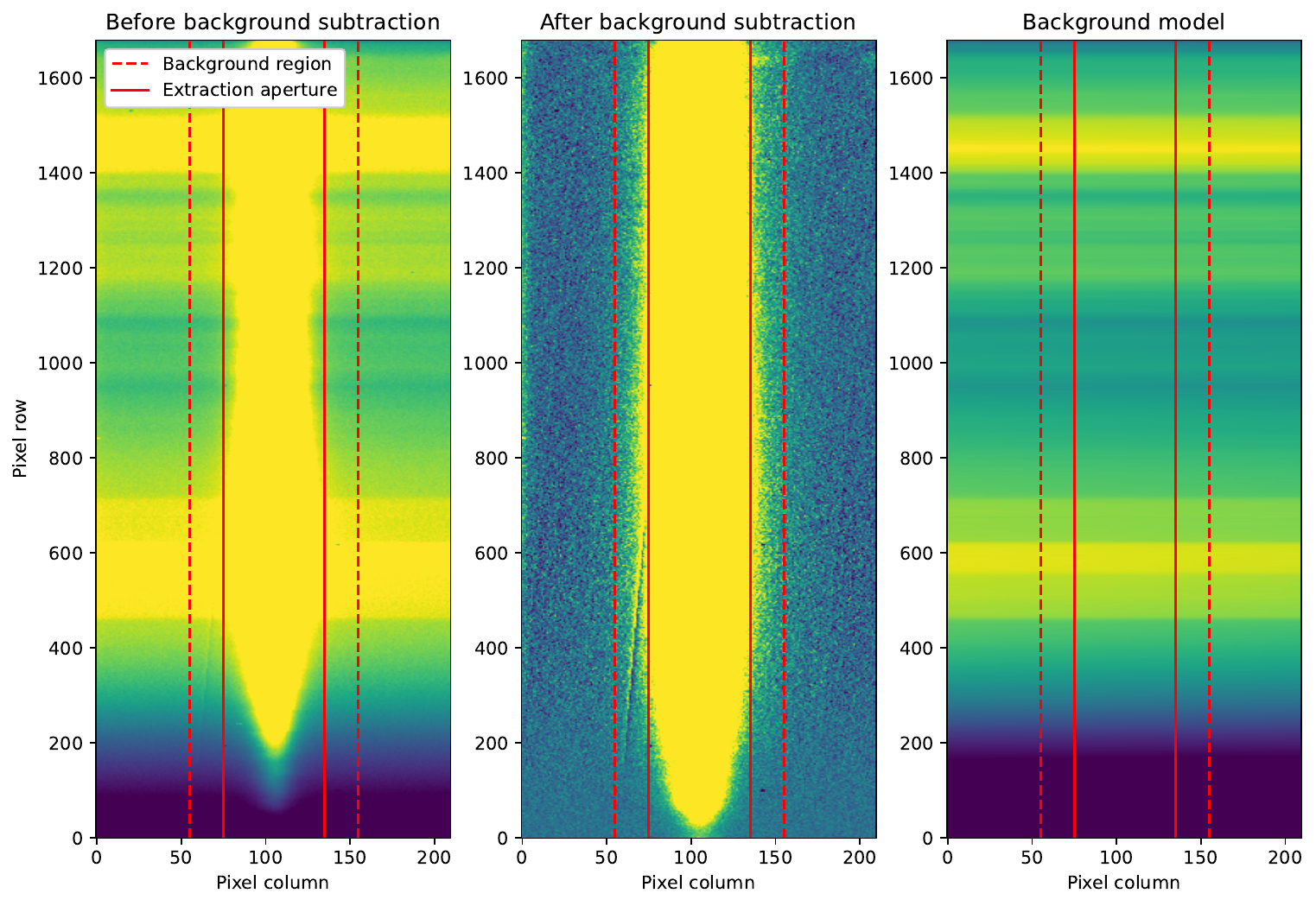}
    \caption{\newerer{Image of the detector region selected for background subtraction in the primary reduction for a single frame of the first night of observation, centred on the fitted trace of the target star (top row) and comparison star (bottom row). The first column shows the region before background subtraction, the second column shows the region after background subtraction, and the fitted linear background model is shown in the third column. The aperture region used for spectral extraction is between the solid red lines, with the background region used for fitting a linear polynomial is without the dotted lines, between which is the a masked offset region. In the post-subtraction image for the target star you can use very faint background stars not visible in the raw image, and in the post-subtraction image of the comparison star you can see a bad pixel column, which is fully contained within the mask for the entire night.}}
    \label{fig:subtraction_primary}
\end{figure*}

\newerer{We use a well-established method of background subtraction following the methodology of previous LRG-BEASTS studies \citep[e.g.,][]{kirk2017rayleigh,alderson2020lrg,ahrer2022lrg}, using a row-by-row polynomial fit within a relatively narrow background region on either side of the aperture. We iterated values for the width of the aperture, background region, and the background offset, and the polynomial order used to fit the background, to minimize the RMS residuals of an initial white-light light-curve fit. We also visually examine the post-subtraction frames and background model to ensure the subtraction method was well-behaved. Our final background subtraction for the first night, as described in Section \ref{sec:reduction}, uses an aperture width of 60 pixels, an offset of 20 pixels, a background region of 55 pixels, and a linear polynomial to fit the background (the background regions are indicated in Fig.\,\ref{fig:extraction-frames}). The pre- and post-subtraction frames, and the fitted background model, for a single frame are depicted in Fig.\,\ref{fig:subtraction_primary}. This background subtraction can be seen to perform well, with no residual sky lines visible. The curvature of the sky lines is negligible (less than one pixel across the regions of interest); if this were not the case we would observe near-horizontal residuals of the sky lines in the post-subtraction image. The diagonal feature in the comparison star frame is a bad column which doesn't extend into the background or source region.
Background subtraction reveals the presence of very faint background stars (too faint to be seen in the raw images) in the post-subtraction background region of the target star, as well as the very tail of the PSF of another background star in the edge of the comparison star background region. Collapsing the flux vertically reveals these stars amount to $\sim0.1\%$ of the total background, resulting in a negligible offset in the transit depth that is constant in time. We also note the poor performance of the background subtraction for y-pixel values above 1600, but highlight that we do not use the flux from these wavelengths in either the white-light or spectroscopic light-curves. The initial white-light light-curve fit for this subtraction (using the same white-light light-curve fitting and detrending described in Section \ref{sec:fitting}) yields a residual RMS of 324 ppm.}

\newerer{To demonstrate the robustness of our spectrum to other choices in background subtraction methodology, we repeat the full host of reduction and analysis steps with alternative choices in subtraction methods, using both the white-light light-curves and resultant transmission spectra to compare between methods. We highlight that three alternative choices: using a wider background region (up to and including the entire detector), using a wider mask to mitigate for any stellar PSF within the background region, and masking the background stars in the primary reduction, all give rise to consistent transmission spectra, and perform equally or worse in terms of residual RMS.}

\begin{figure*}
    \centering
    \includegraphics[width=0.9\textwidth]{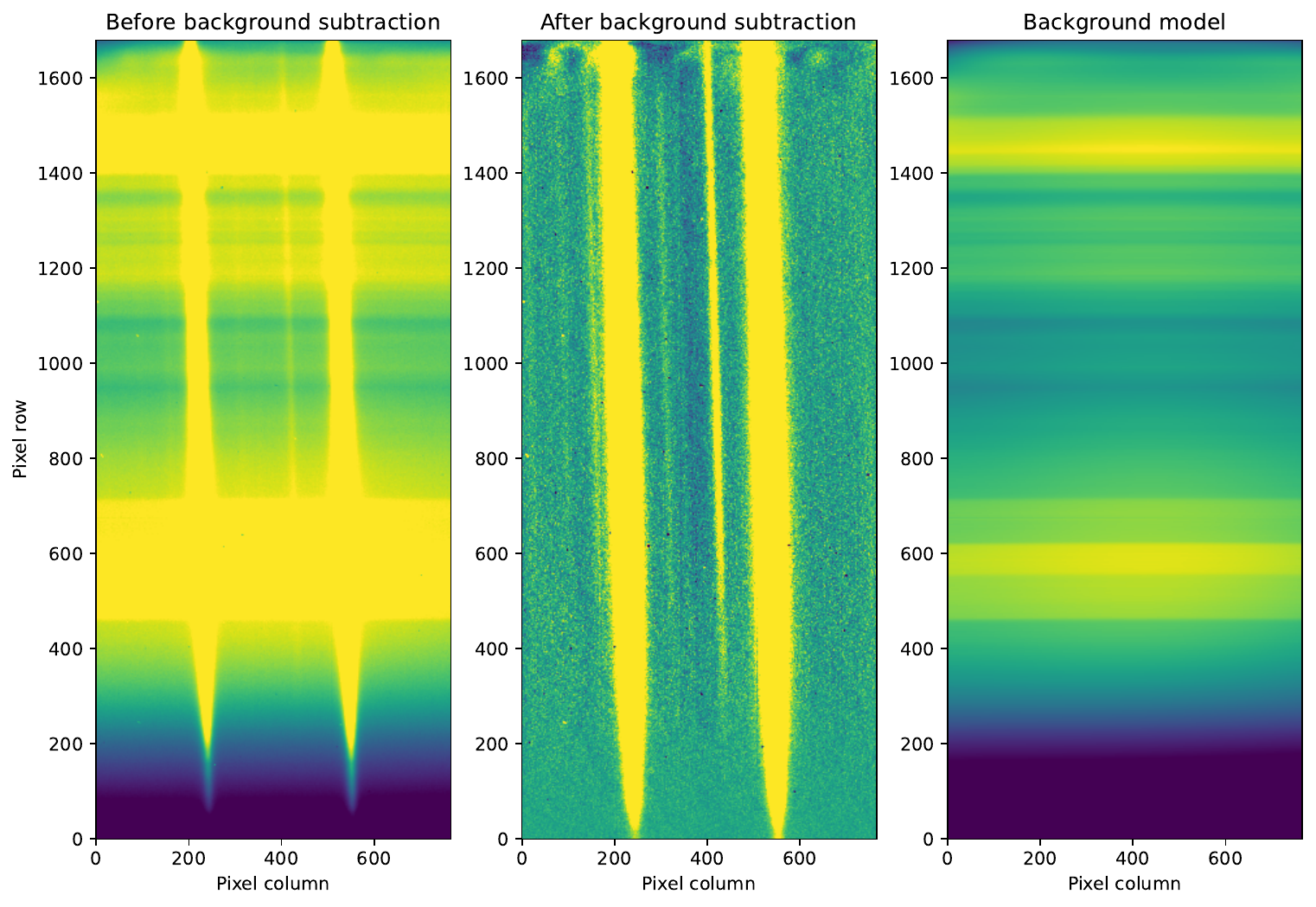}
    \caption{\newerer{Image of the entire detector selected for background subtraction in an alternative reduction for a single frame of the first night of observation. The first column shows the region before background subtraction, the second column shows the region after background subtraction, and the fitted quartic background model is shown in the third column.}}
    \label{fig:subtraction_full_detector}
\end{figure*}

\newerer{The level of the background far from the source is uninformative on the level of the background within the aperture because the background varies in the cross-dispersion direction in a relatively complicated manner. This motivates the usage of a relatively narrow background region, where the background level has a simpler functional form. This typically means that a higher-order polynomial is required when a wider background region is used. This is confirmed when we test wider backgrounds regions, for example with a wider region of 100 pixels, the subtraction fares poorly with a linear fit (white-light light-curve RMS of 442 ppm, 188 ppm worse than our primary reduction), but improves notably with a quadratic fit (white-light light-curve RMS of 348 ppm, 24 ppm worse). When using a background fit to the entire detector (excluding 10 pixels at each edge and masking both stars), this behaviour holds, with linear, quadratic, and cubic fits all performing very badly. With a quartic fit to the entire detector, we are able to approach the performance of the primary subtraction, with a white-light light-curve RMS of 370 ppm, 46 ppm worse than the primary reduction. We present the pre- and post-subtraction frames and fitted background model with this method in Fig.\,\ref{fig:subtraction_full_detector}. We highlight again the lack of any diagonal features from curvature of the sky lines, and the presence of the same background stars in the post-subtraction frame that appeared in the primary subtraction.}

\newerer{To verify that starlight from the wings of the PSF bleeding into the background region is not affecting the results, we repeat the background subtraction with a broader background offset of 30 pixels and reduce the background width to 45 pixels (so no new pixels are considered). This has virtually no impact on the post-subtraction data, with a white-light light-curve fit residual RMS of 323 ppm (a 1 ppm improvement). We note that the white-light $R_p/R_*$ is effectively identical as well (72 ppm deeper, compared to an error of 1300 ppm), demonstrating that the choice of a wider mask has a negligible impact on the transit depth. We also repeat our primary reduction but masking the pixels including the background stars centered at 67 pixels to the left of, and 85 pixels to the right of, the target star. These stars are very faint (about 35 and 20 counts at their peak pixels respectively), and masking them has negligible impact again, with an unchanged white-light light-curve residual of 324 ppm, and a change in $R_p/R_*$ of only 138 ppm shallower.}

\begin{figure*}
    \centering
    \includegraphics[width=0.97\textwidth]{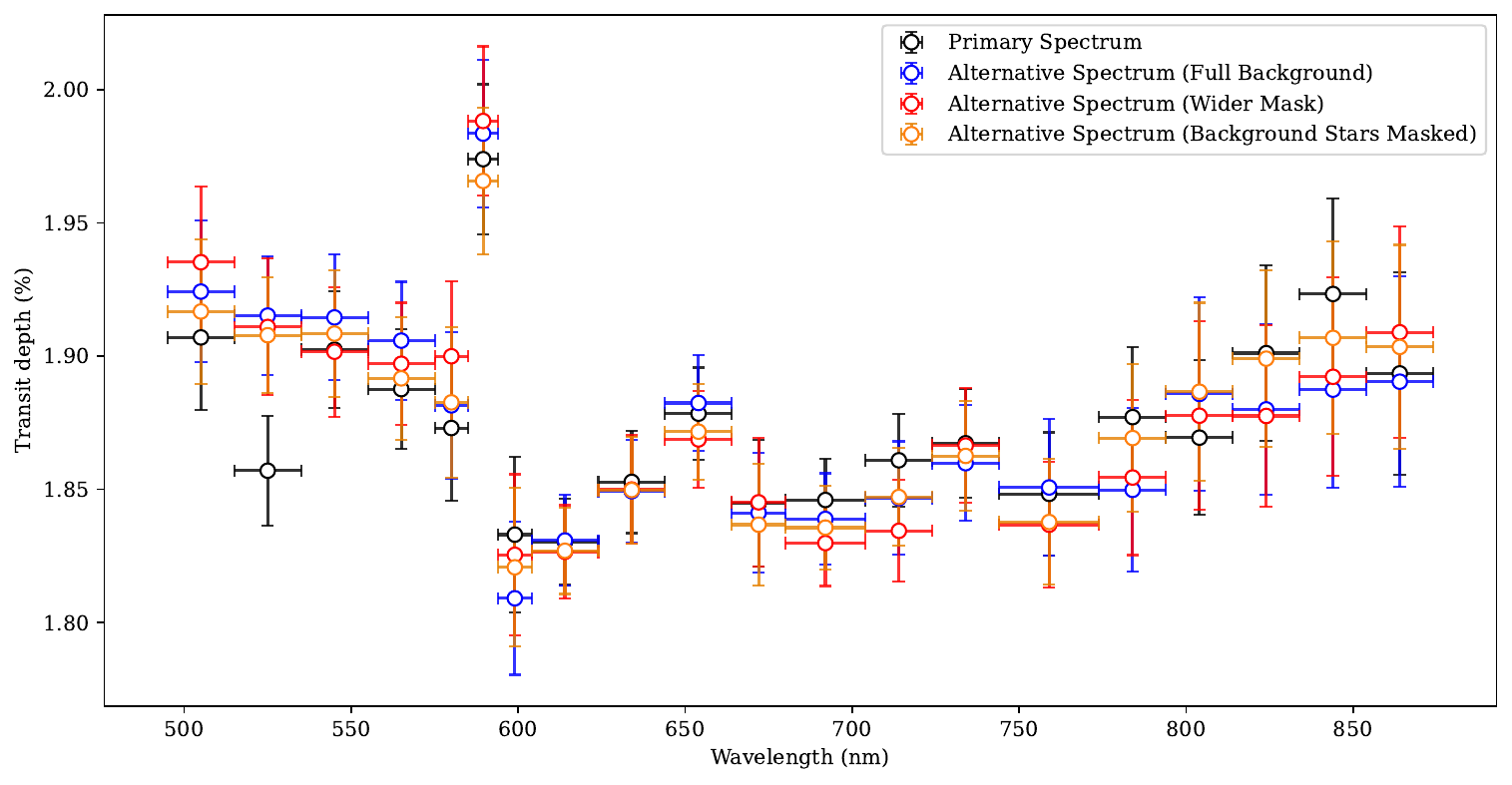}
    \caption{\newerer{Transmission spectrum from the primary data reduction from the first night (black data points), compared to alternative methods using different background subtraction techniques. The same spectral features are present in each of the reductions, including sodium absorption, and the red-ward and blue-ward slopes associated with \ch{CH4} and \ch{H2O} absorption, and rayleigh-scattering respectively.}}
    \label{fig:spectra_subtractions}
\end{figure*}

\newerer{All of these different background subtraction methods produce highly concordant transmission spectra, presented in Fig.\,\ref{fig:spectra_subtractions}, demonstrating that our results are robust to all of these decisions. Therefore, the interpretations of the paper, including sodium absorption and the presence of blue-ward and red-ward slopes, are not sensitive to the specific choices made in the background subtraction. We note that the use of the whole detector gives rise to more noise in both white-light and spectroscopic light-curves and requires a higher order polynomial to fit the background, confirming that the background flux far from the trace is not helpful in determining the shape of the background within the aperture.}

\section{Red noise analysis}

\begin{figure*}
    \centering
    \includegraphics[width=0.97\textwidth]{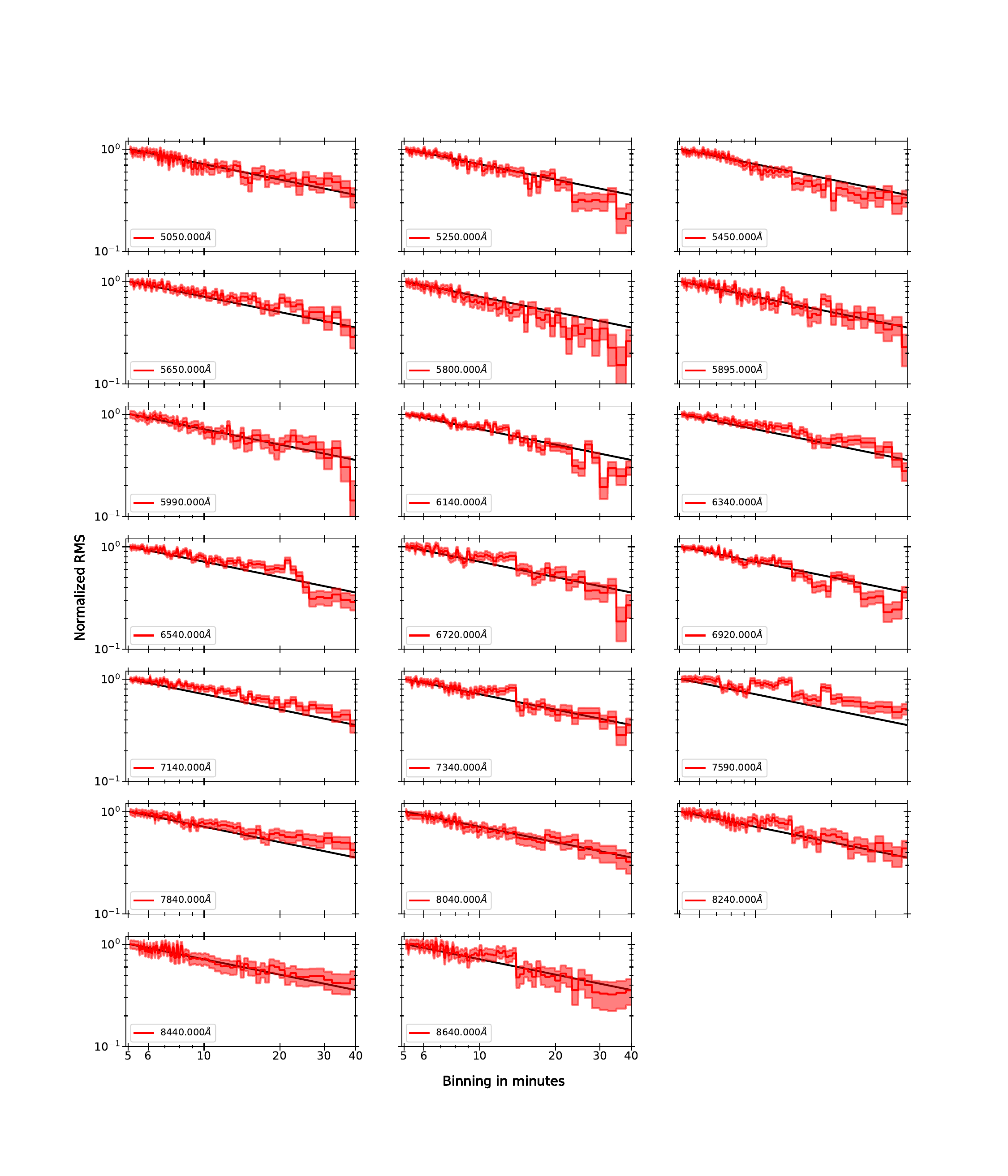}
    \caption{\newest{Root mean square of the spectroscopic light-curve fit residuals from the first night as a function of binning in time, compared to the $1/\sqrt{N}$ expected from white noise. For each wavelength bin the measured noise is consistent with the uncertainties scaled to the binning, which are $\sim$ 200\,ppm at 10 points per bin. The single exception to this is the wavelength bin centered at 7590 \AA, which contains the telluric \ch{O2} absorption feature. This single bin does not drive our interpretation of the transmission spectrum.}}
    \label{fig:rms_vs_bins}
\end{figure*}

\newest{To demonstrate the minimal impact of red noise on our transmission spectrum, we present plots of the root mean square residuals of the spectroscopic light-curve fits from the first night of observation in Fig.\,\ref{fig:rms_vs_bins}.}

\section{All atmospheric models}

\begin{figure*}
    \centering
    \includegraphics[width=0.97\textwidth]{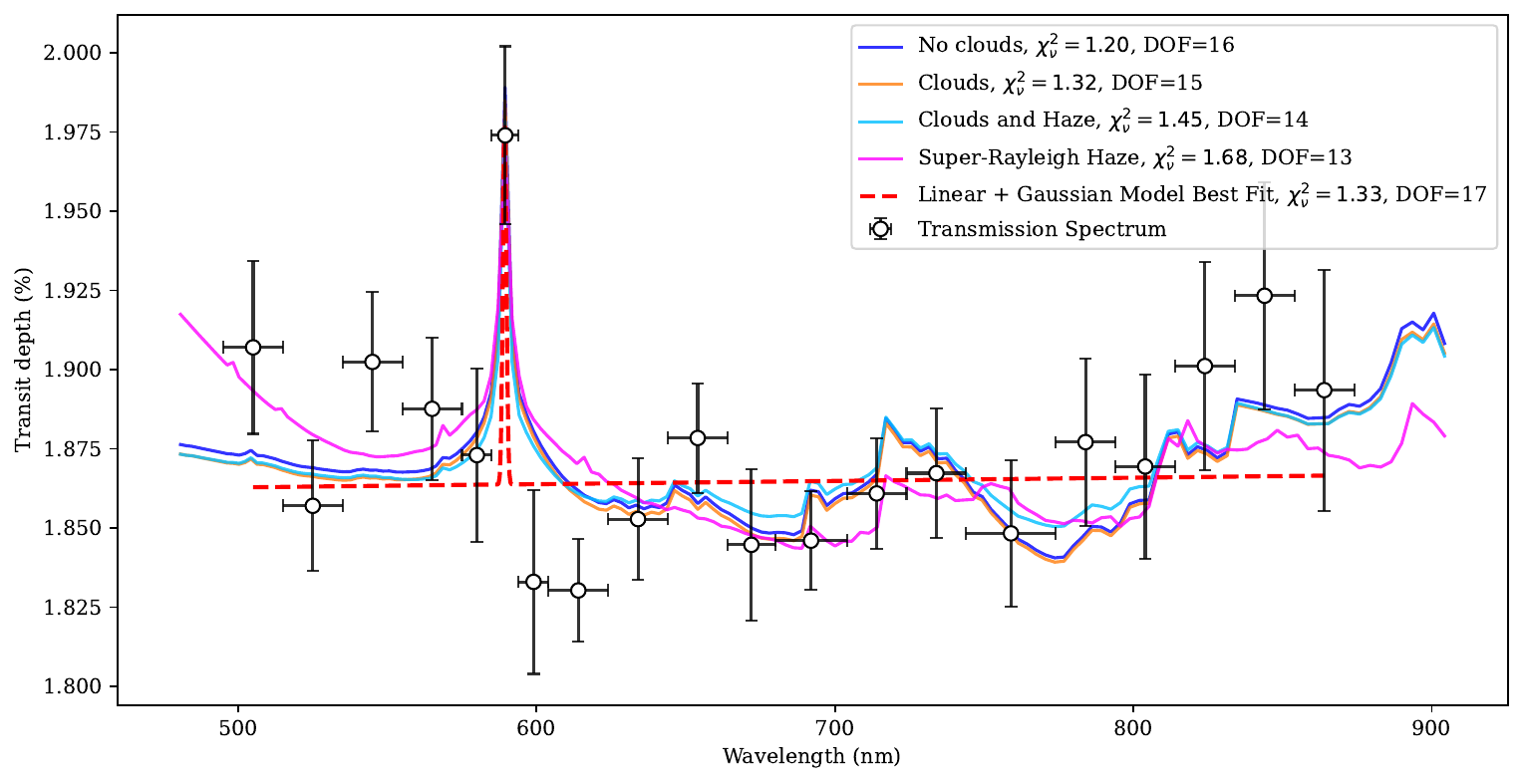}
    \caption{\newest{Transmission spectrum from the first night with (black data points) and the full set of best-fitting {\tt petitRADTRANS} model from our atmospheric retrieval analysis, plus our best-fitting simple parametric fit (red dotted line). The plotted retrieval fits include a model with no aerosols (dark blue), a model with free grey cloud-top pressure (orange), a model with free cloud-top pressure and a free enhancement of the Rayleigh scattering slope representing haze particles (light blue), and a model with free cloud-top pressure and a free (permitting super-Rayleigh) scattering slope (magenta).}}
    \label{fig:spectrum_all_fits}
\end{figure*}

\newest{We present the full set of fits to the transmission spectrum from the \texttt{petitRADTRANS} atmospheric retrievals in Fig.\,\ref{fig:spectrum_all_fits}, in addition to the best simple parametric fit.}

\bsp	
\label{lastpage}
\end{document}